\documentclass[11pt,a4paper]{article}

\usepackage{jheppub}
\usepackage{graphicx}
\usepackage{amsmath,amsfonts,amssymb}
\usepackage{wrapfig}
\usepackage{hyperref}
\usepackage{color}
\usepackage{verbatim}
\usepackage{physics}
\usepackage{cancel}
\usepackage{enumerate}

\def\lsim{\mathrel{\raise.3ex\hbox{$<$\kern-.75em\lower1ex\hbox{$\sim$}}}}
\def\gsim{\mathrel{\raise.3ex\hbox{$>$\kern-.75em\lower1ex\hbox{$\sim$}}}}

\newcommand{\be}{\begin{equation}}
\newcommand{\ee}{\end{equation}}
\newcommand{\bea}{\begin{equation}\begin{aligned}}
\newcommand{\eea}{\end{aligned}\end{equation}}

\newcommand{\muf}{\mu_0}

\newcommand{\td}[1]{\tilde{#1}}

\definecolor{lightblue}{RGB}{102,204,255}

\title{An EFT approach to the study of multi-phase criticality scenarios}

\author[a]{Kristjan Kannike,}
\author[a,b]{Luca Marzola,}
\author[a,c]{Kristjan M\"u\"ursepp}

\affiliation[a]{National Institute of Chemical Physics and Biophysics, R\"{a}vala 10, 10143 Tallinn, Estonia}
\affiliation[b]{Institute of Computer Science, University of Tartu, Narva mnt 18,
51009 Tartu, Estonia}
\affiliation[c]{INFN, Laboratori Nazionali di Frascati, C.P. 13, 100044 Frascati, Italy}
\emailAdd{kristjan.kannike@cern.ch}
\emailAdd{luca.marzola@cern.ch}
\emailAdd{kristjan.muursepp@kbfi.ee}

\abstract{
Multi-phase critical scenarios explain the observed Higgs boson mass scale by the almost simultaneous occurrence of two smoothly connected phases of the theory, which differ by the selected vacuum configuration. A generic prediction of the framework is the presence of a further light scalar state, the dilaton, which naturally couples weakly to the Higgs boson. The implementation of the framework usually requires the presence of a third, heavier state, which plays the role of dark matter and ensures the couplings run so that the multi-phase criticality condition is met. In this paper we consider the multi-phase criticality limit of an extension of the Standard Model including two extra scalar singlets, addressing the scenario with effective field theory methods that are particularly suited for treating the hierarchical mass spectrum that this construction yields. The analysis improves on the approximated results available in the Literature and explores the phenomenology of the model at collider and dark matter experiments. We find that the running of scalar couplings in the EFT between the two scales cannot be ignored, but the quantum corrections from the dark matter candidate are not noticeably modified.
}

\begin{document}

\maketitle

\section{Introduction}
\label{sec:intro}

In scenarios where novel particles  
interact with the fields of the Standard Model (SM), the parameters of the latter receive contributions from new quantum corrections that are proportional to the energy scale at which the new degrees of freedom emerge. However, the null results obtained from new physics searches at current and past collider experiments have already pushed the lowest energy at which new effects could manifest well above the electroweak scale. Consequently, in the absence of symmetries that shield the SM from such contributions, parameters like the Higgs boson mass are progressively driven away from natural values close to the electroweak scale, thereby posing the basis for the so-called hierarchy problem. 

This shortcoming, grounded in our present understanding of quantum corrections, has motivated the proposal of several theoretical constructions where these contributions are naturally suppressed. Among the most prominent examples is the mechanism of dynamical symmetry breaking, first proposed by Coleman and Weinberg (CW)~\cite{Coleman:1973jx}. In this case, the tree-level scalar potential is endowed with an additional scale symmetry that quantum corrections explicitly break to yield the non-vanishing vacuum expectation value (VEV) acquired by the scalar field in the theory.

The original CW construction can be generalised to include multiple scalar fields, provided that their potential possesses a flat direction, as first noticed by Gildener and Weinberg (GW)~\cite{Gildener:1976ih}. In the GW framework, the tree-level potential admits a whole line of degenerate minima in the field space, the flat direction, along which quantum corrections dominate the dynamics. Depending on specific mass scales---and, in turn, on particular field values---these corrections lift the degeneracy of the potential, determining a minimum whose radial position is closely related to that of the remnant of the imposed tree-level flat direction. As a result, the scalar fields of the theory develop non-vanishing VEVs, which guarantee the presence of at least one naturally light state: the \textit{dilaton}, the pseudo-Nambu-Goldstone boson of the spontaneously broken scale invariance. 

Whereas identifying the Higgs boson with the dilaton would certainly be a way to solve the hierarchy problem, the present experimental results constraining the parameters and couplings of the Higgs boson disfavour this possibility. Nevertheless, the idea could still be partially salvaged if the Higgs boson were to be associated with a pseudo-Nambu-Goldstone boson that emerges from the breaking of a custodial global symmetry at some high scale~\cite{deBoer:2024jne, deBoer:2025oyx}. This elegant mechanism, known as custodial naturalness, addresses the hierarchy problem via the running of the scalar couplings, by imposing specific boundary conditions at a high scale that conform to the assumption of custodial symmetry. 

Another possibility is that the Higgs boson and the dilaton acquire their mass in a multi-phase critical (MPC) scenario~\cite{Kannike:2021iyh}. In this case, near the critical boundary where two phases coexist, the Higgs mass remains naturally light provided that it changes sign while transitioning between the adjacent phases. Similarly, the mixing angle between the dilaton and the Higgs field is also naturally suppressed near the same critical boundary, as required by the present collider bounds~\cite{Huitu:2022fcw}. 

The MPC scenario can be realised by augmenting the SM with two scalar field singlets, resulting in a simple setup that nevertheless is able to connect the electroweak symmetry breaking (EWSB) dynamics to dark matter (DM) phenomenology~\cite{Kannike:2022pva}. MPC is then realised by choosing a suitable hierarchy between the couplings of the model,  
related to the hierarchy of the field-dependent masses and rendering the symmetry-breaking mechanism qualitatively different from the usual GW construction. In particular, the renormalisation group (RG) evolution of couplings between the flat direction and the minimum scale has to be taken into account in the computation of the mass eigenstates of the theory \cite{Alexander-Nunneley:2010tyr,Kannike:2020ppf}. This yields implications for the Higgs-dilaton coupling probed at collider, as well as for DM phenomenology~\cite{Kannike:2020ppf, Kannike:2022pva}. In this regard, the properties of DM within a two-singlet-scalar extension of the SM were explored in~\cite{Abada:2011qb,Abada:2012hf,Abada:2013pca}, in the non-scale invariant case and in~\cite{Ishiwata:2011aa,Gabrielli:2013hma,Kang:2020jeg} in the scale invariant limit. However, these studies focused only on the EWSB dynamics and did not analyze the specific signatures entailed by the MPC scenario. For instance, it has been argued that the two-singlet model in the MPC regime may result in an observable gravitational wave signal emitted in a first-order phase transitions or in the production of primordial black holes ~\cite{Arteaga:2024vde}.

In this work, to exhaustively investigate the MPC construction and go beyond the results of~\cite{Kannike:2022pva}, we adopt the two-singlet extension of the SM and investigate the framework paying greater attention to the problem posed by the presence of different mass scales. In fact, a viable implementation of the MPC scenario requires the presence of a state, our DM candidate, that is much heavier than the Higgs boson and the dilaton. Several characteristic mass scales are then present in the broken phase of the theory, potentially leading to significant uncertainties in the calculation of the scalar potential. Addressing these uncertainties requires the use of RG-improvement techniques~\cite{Chishtie:2006nr, Manohar:2020nzp} needed to bring the logarithmic corrections stemming from higher-order contributions under control. Moreover, in this improved analysis, we also investigate the magnitude of the previously neglected one-loop corrections from the top quark and the SM gauge bosons, as well as those from the Higgs and the dilaton.

For a global analysis of the scalar potential of one field, or along a particular direction in the scalar field space, the most common approach to RG improvement sets the RG scale, $\mu$, to a field-dependent value, $\mu = m({\phi})$, where $m(\phi)$ is the field-dependent mass of the relevant scalar field $\phi$. This choice is equivalent to re-summing all the leading logarithms in the scalar potential provided that the coupling constants are evaluated at the field-dependent scale $m(\phi)$ using one-loop RG equations (RGEs). However, for very restricted regions of field space, a field-dependent renormalisation scale may often be an impractical choice, especially if the RGEs do not admit analytical solutions. In this case one may instead employ a fixed scale RG-improvement method, for instance by approximately re-summing the leading logarithms by a set mass $\mu= m(v)$, corresponding to the physical mass of the scalar field. This requires an approximate knowledge of the expectation value $\expval{\phi} = v$ of $\phi$, and allows to study the potential only around the corresponding minimum in field space.

A further complication arises in theories with multiple mass scales, wherein a single choice of the scale $\mu$ cannot even approximately guarantee the re-summation of all the logarithms from loop corrections as those necessarily involve different mass scales. For example, consider two scalars $\phi_1$ and $\phi_2$, with VEVs $v_1$ and $v_2$, and masses $m_1(\phi_1, \phi_2) \ll m_2(\phi_1, \phi_2)$. In this case, setting $\mu = m_1(v_1,v_2)$ allows to approximately re-sum the leading logarithms depending on $m_1$ around the position of the VEV in field space, but the choice introduces large corrections of the form $\ln \left(m_2(v_1,v_2)/m_1(v_1,v_2) \right)$ which hinder the validity of the underlying perturbative approach. More sophisticated methods are thus needed.

One possible approach to treat multi-scale potentials is to introduce multiple renormalisation scales, as discussed in~\cite{EINHORN1984261, Ford:1996hd, Ford:1996yc, Steele:2014dsa}. However, as emphasised in~\cite{Manohar:2020nzp}, this method is difficult to put into practice and still does not allow to re-sum all the higher-order corrections using the RGEs alone. Alternatively, problems involving several characteristic scales may be tackled with methods of Effective Field Theory (EFT)~\cite{Bando:1992wy, Casas:1998cf,Iso:2018aoa, Okane:2019npj, Manohar:2020nzp}, by constructing a tower of theories that can be treated with single-scale techniques and that, together, cover the energy span of the original framework. An excellent account of RGE-improvement methods for multi-scale potentials can be found in~\cite{Manohar:2020nzp}. 

For the case of the MCP scenario, broadly speaking, the EFT procedure may be summarised in three steps. First, one computes the contribution to the CW potential due to the heavy field that plays the role of DM, evaluating it at the corresponding large mass in the RG-improvement procedure. This allows to re-sum the leading logarithmic contributions involving the heavy degree of freedom. After that, the high-energy theory can be matched onto an EFT obtained by integrating out the same heavy degree of freedom. Finally, the EFT couplings are run down to the scale of the light degrees of freedom, the Higgs boson and the dilaton, whose contributions to the CW potential can now be evaluated at the corresponding characteristic energy scale. Provided that one also takes proper care of tadpole conditions and shifts the scalar fields by their background values before constructing the EFT, the resulting potential will be properly RG-improved and thus presents no large logarithmic corrections. The procedure can be straightforwardly generalised if more characteristic scales are considered in the theory, for instance, if the masses of the Higgs boson and the dilaton were to show a sizeable hierarchy. Detailing this construction and exploring its full implications for the MPC scenario is the main objective of the present analysis. 

The paper is organised as follows. In Section \ref{sec:eft:gw} we provide a short overview of the computation of the effective potential, of the GW construction and also briefly review how multi-scale potentials can be RG-improved by using EFT methods. Next, in Section~\ref{sec:model}, we introduce the MPC scenario and explain how it differs from the standard GW case. We highlight these differences in a specific 2-singlet extension of the SM, as it provides the simplest implementation of the MPC framework. In Section~\ref{sec:EFT_treatment} we extend the previous discussion by addressing the model within the rigorous framework of EFT for its RG-improvement, focusing on prototypical cases with different mass spectra. The results and the phenomenological implications of the model on DM physics are presented in Section \ref{sec:res}. Our conclusions and comments are outlined in Section \ref{sec:concl}. 

\section{Effective potential}
\label{sec:eft:gw}

To find the true pattern of spontaneous symmetry breaking, we have to calculate the effective potential which also takes into account quantum effects. The effective potential is generically given as an expansion
\begin{equation}
  V =\sum_{i=0}^\infty V^{(i)},
\label{eq:V:eff}
\end{equation}
where $V^{(0)}$ is the tree-level contribution and $V^{(i)}$ is the $i$th loop correction. By using dimensional regularisation in the $\overline{\text{MS}}$ scheme, the one-loop contribution is given by
\begin{equation}
    V^{(1)} = \frac{1}{64 \pi^{2}} \sum (-1)^{2 s_i} n_i m_i^{4} 
    \left(\ln \frac{m_i^{2}}{\mu^{2}} - c_i \right),
\label{eq:V:(1)}
\end{equation}
where $m_i$ are the field-dependent mass eigenvalues, $n_i$ are the degrees of freedom, $\mu$ is the renormalisation scale, and $c_{i} = \frac{3}{2}$ for scalars and fermions, whereas $c_{i} = \frac{5}{6}$ for gauge bosons. The sum is taken over all the degrees of freedom present in the theory, including Goldstone bosons. 

Requiring that the effective potential in Eq.~\eqref{eq:V:eff} remain invariant under changes of the (arbitrary) renormalisation scale $\mu$ determines the RG evolution of the couplings of the theory, as dictated by the Callan-Symanzik equations. This improvement procedure then effectively replaces the arbitrary $\mu$ with a scale $\mu_0$ at which the values of the couplings must be set. Then if all degrees of freedom of the theory have comparable masses, $m_i \simeq \tilde{m}$, this procedure is sufficient to ensure that the logarithmic corrections 
remain well under control provided that a scale $\mu_0 \simeq \tilde{m}$ is chosen. To the contrary, hierarchical mass spectra yield corrections sourced by the large mass ratios that require further treatment, as briefly reviewed in Sec.~\ref{sec:multiscale}. Having said that, for the moment, we focus on the simple case where the improvement of the effective potential is enough to avoid problems with the re-summation of these large logarithmic contributions.    

We remark that the one-loop correction~\eqref{eq:V:(1)} can also be rewritten in the form 
\begin{equation}
    V^{(1)} = A \varphi^4 + B \varphi^4 \ln \frac{\varphi^{2}}{\mu^{2}},
\label{eq:V:(1):A:B}
\end{equation}
where
\begin{equation}
  A = \frac{1}{64 \pi^{2} v_\varphi^4} \sum (-1)^{2 s_i} n_i m_i^{4} 
    \left(\ln \frac{m_i^{2}}{v_\varphi^{2}} - c_i \right),
    \qquad
 B = \frac{1}{64 \pi^{2} v_\varphi^4} \sum_i (-1)^{2 s_i} n_i m_i^{4},
\label{eq:AB}
\end{equation}
and $\varphi$ denotes the radial direction in field space along which the VEV $v_\varphi$ is located.

\subsection{Gildener-Weinberg method}
\label{subsec:GW}
Usually the tree-level contribution is much larger than the one-loop contribution. The GW construction \cite{Gildener:1976ih} ensures that quantum corrections come to dominate the potential in a region of field space where the tree-level contribution is set to vanish, thereby inducing non-vanishing VEVs for the scalar fields involved by breaking scale symmetry. To illustrate the scenario, consider the simplest case of a scale-invariant scalar potential of two real fields $\phi$ and $\chi$, given at the tree-level by
\be
\label{v0gww}
V^{(0)}= \frac{1}{4}\lambda_\chi \chi^4 + \frac{1}{4}\lambda_{\chi \phi} \chi^2 \phi^2 + \frac{1}{4}\phi^4.
\ee
The potential has a flat direction in field space along which it vanishes, $V^{(0)} = 0$,  if the couplings satisfy the condition
\begin{equation}
\label{ffuc}
\lambda_{\chi\phi} = -2 \sqrt{\lambda_\chi \lambda_\phi}.
\end{equation}
The flat direction lies along the direction $\hat n$ tilted with respect to the $\chi$-axis of field space by an angle $\theta$ determined by
\begin{equation}
    \tan^2 \theta = \sqrt{\frac{\lambda_\chi}{\lambda_\phi}}\,,
    \label{eq:flatdir}
\end{equation}
and $V^{(0)}$ has then non-vanishing curvature only in a direction orthogonal to $\hat n$. The field-dependent mass matrix has two eigenvalues $m_1$ and $m_2$. The mode associated to the flat direction, the dilaton, remains massless in first approximation ($m_1=0$ at tree level). Then, denoting the non-zero  field-dependent mass by $m_2$, the one-loop correction to the potential reads
\begin{equation}
    V^{(1)} = \frac{1}{64 \pi^2} m_2^4 \left(\frac{m_2^2}{\mu^2} - \frac{3}{2}\right)\,.
\label{eq:V:1:toy}
\end{equation}
 The dependence on the arbitrary renormalisation scale $\mu$ is addressed by the RG-improvement procedure, which introduces a specific scale $\muf$ in its place. As a result, the condition in Eq.~\eqref{ffuc} now holds only at the scale $\muf$, which can be identified with the flat-direction scale, where the running of couplings is initialised. The GW potential along the flat direction can be rewritten as 
\begin{equation}
    V = B \varphi^4 \left(\frac{\varphi^2}{v_\varphi^2} - \frac{1}{2} \right)\,,
\end{equation}
with $B = m_2^4/(64 \pi^2 v_\varphi^4)$ and $\varphi$ the radial scalar mode along $\hat n$---the dilaton. The dilaton VEV $v_\varphi$ is solely determined by quantum corrections along the flat direction and sets the VEVs of the original scalar fields as $\ev{\chi} = v_{\varphi} \cos \theta$ and $\ev{\phi} = v_{\varphi} \sin \theta$. Since the scale symmetry is explicitly broken by the loop corrections, $\varphi$ also obtains a non-zero mass
\begin{equation}
    M_1^2 = 8 B v_\varphi^2 = \frac{m_2^4}{8 \pi^2 v_\varphi^2},
    \label{eq: LightMass}
\end{equation}
while the correction to the mass of the orthogonal direction is negligible: $M_2 \approx m_2$. In the GW construction, the mass eigenstates corresponding to the $M_{1}$ and $M_2$ eigenvalues lie in field space along the flat direction and orthogonally to it, respectively. The mixing of the scalar fields in Eq.~\eqref{v0gww} then coincides with the angle giving the tilt of the flat direction. Another relation peculiar to this framework is that between the flat direction scale and the dilaton VEV
\begin{equation}
    \mu_0 = v_\varphi\, e^{\frac{1}{4} + \frac{A}{2 B}},
\end{equation}
which, for the potential Eq.~\eqref{eq:V:1:toy}, evaluates to
\begin{equation}
    \mu_0 = \frac{m_2}{\sqrt{e}}\,.
\end{equation}


\subsection{EFT improvement of multiscale potentials}
\label{sec:multiscale}
Previously we discussed how choosing an appropriate scale $\mu_0 \simeq \tilde{m}$, and evaluating all the couplings at $\mu_0$ can limit the impact of the logarithms due to loop corrections, thereby ensuring the validity of the perturbative expansion in a region of field space close to the minimum of the potential. However, the argument above fails when different mass scales are present in the theory. To illustrate this, consider a scenario with two hierarchical mass scales $m_1 \ll m_2$. The one-loop potential is then schematically given by
\begin{equation}
    V^{(1)} = \frac{m_1^4}{64 \pi^2} \left( \ln  \frac{m_1^2}{\mu^2} -\frac{3}{2} \right) + \frac{m_2^4}{64 \pi^2} \left( \ln  \frac{m_2^2}{\mu^2} -\frac{3}{2} \right)\,.
\end{equation}
Then, improving the potential by setting the scale as $\mu_0 \simeq m_1$ or $\mu_0 \simeq m_2$, does not prevent potentially large logarithms of the form $\pm \ln \left(m_2^2/m_1^2\right)$ from appearing in the theory. 

An elegant way to avoid this apparent complication is to properly treat the heavy degrees of freedom, which should decouple from the theory at low energies. This can be naturally achieved using the methods of EFT, as described comprehensively in~\cite{Manohar:2020nzp}. 

The first step of this procedure is to initialise the couplings of the theory at a heavy mass scale $\mu_H$, perform a shift in the path integral by the constant background field values and then integrate out the heavy fields. After this step, one constructs an EFT valid below the scale $\mu_H$ by matching the cosmological constant and the $N$-point amplitudes of the low energy effective theory to the high energy theory amplitudes, at one-loop and tree-level, respectively. As a next step, the couplings of the EFT are run down from $\mu_H$ to a low energy scale $\mu_L$ at which only the IR degrees of freedom propagate. These light fields are then integrated out, and the next-to-leading logarithmic (NLL) contributions to the effective potential can be evaluated. 

The scalar potential resulting from this procedure can be thought of as the sum of 1PI vacuum diagrams with vanishing tadpole conditions \cite{PhysRevD.9.1686}. This means that the classical source terms $J_{\phi}$ have to be chosen in such a way that $\expval{\phi} = 0$ for all the fields $\phi$ appearing in the path integral up to the one-loop level. If $J_{\phi}$ contains large logarithms, they have to be RG-improved as well.

Another subtlety of the EFT approach, clearly explained in~\cite{Manohar:2020nzp}, concerns the non-commutativity of shifting the fields in the path integral by their background field value and constructing the EFT. This is because, in the EFT, the information about 1PI diagrams is obfuscated by the fact that the heavy particle lines have been shrunk. Thus, as long as the matching between the EFT and the high energy theory is non-trivial, the shift in the path integral should always be performed in the high energy theory and not in the EFT. We will pay special attention to this issue when studying the RGE improvement of potentials exhibiting MPC, with the exception of scenarios wherein the matching is trivial and, thus, the fields in the path integral can be safely shifted by their background field values in the EFT.

\section{Multi-phase criticality}
\label{sec:model}
Before going into the technicalities of the EFT approach, let us first review the MPC framework using the model that we will study in detail later, obtained by extending the SM with two singlet scalar fields enjoying a $\mathbb{Z}_2$ symmetry. The scalar sector of the theory presents one Higgs doublet $H = \qty(G^+, (h+ i G^0)/\sqrt{2})^T$, comprising the Higgs boson, $h$, and two Goldstone bosons, $G^{0}$ and $G^{+}$, as well as two scalar singlets, the dilaton-like $s$ and the DM candidate $s'$:
\begin{equation}
\begin{split}
    -\mathcal{L} &\supset \lambda_h \abs{H}^4 + \frac{1}{4}\lambda_s s^4  + \frac{1}{4}\lambda_{s'}s'^4 + \frac{1}{2}\lambda_{hs} \abs{H}^2 s^2 +  \frac{1}{2}\lambda_{hs'}\abs{H}^2 s'^2  + \frac{1}{4}\lambda_{ss'} s^2s'^2 \,. 
    \label{eq:LUV}
\end{split}
\end{equation} 
The potential above respects a $\mathbb{Z}_2 \times \mathbb{Z}_2$ symmetry, under which the singlets transform as $s \to -s$ and $s' \to -s'$, respectively.\footnote{Notice that it is enough to impose a $\mathbb{Z}_2$ symmetry on $s'$, after which an accidental $\mathbb{Z}_2$ symmetry appears in the potential for $s$ as well.} We choose the quartic couplings so that $s'$ never develops a non-zero VEV: the $s'$ self-coupling and its portal couplings then must not acquire negative values at any scale. The field $s'$ can thus be a DM candidate, and its couplings to $h$ and $s$ will be the source of the quantum corrections.

The Lagrangian in Eq.~\eqref{eq:LUV} then supports the existence of the following phases:

\begin{enumerate}[i)]
    \item $\expval{s}\neq 0$ and $\expval{h} = 0$, realised as the critical boundary $\lambda_s=0$ is crossed. For this to happen dynamically, the positivity of the potential implies that the RG evolution of the parameter must drive $\lambda_s$  to progressively larger values---its $\beta$-function must be positive, $\beta_{\lambda_s}>0$. A further condition, $\lambda_{hs}>0$, ensures that the tree-level Higgs mass is positive and consequently prevents the occurrence of mixing in the scalar sector formed by $h$ and $s$.  
    \item $\expval{h}\neq 0$ and $\expval{s} = 0$, complementary to the previous case \cite{Chway:2013fzr}. In this phase, the running of $\lambda_h$ is responsible for dynamically taking the parameter through the $\lambda_h=0$ boundary. The condition $\lambda_{hs}>0$ ensures the positivity of the $s$ scalar mass and prevents the emergence of mixing. This possibility, which motivated the original GW proposal, could be disproved by future measurements of the Higgs boson trilinear coupling.
    \item $\expval{h}\neq 0$ and $\expval{s} \neq 0$. A phase where both the scalars develop non-vanishing VEVs appears as the critical boundary
    \begin{equation}
    \lambda_{hs} + 2\sqrt{\lambda_h\lambda_s}=0
    \label{eq:shcritb}
    \end{equation}
    is crossed. The positivity of the potential further requires $\lambda_h\geq0$,  $\lambda_s\geq0$ and, consequently, $\lambda_{hs}<0$. In a pure CW construction, the scalar field VEVs would single out a point on a flat direction tilted at an angle $\expval{s}/\expval{h} = (\lambda_h/\lambda_s)^{1/4}$ (see e.g. \cite{Hempfling:1996ht,Foot:2007iy}). Dynamical symmetry breaking then takes place if the evolution of the couplings in Eq.~\eqref{eq:shcritb} pushes  the effective coupling regulating the radial profile of the potential along the tree-level flat direction to positive values.
\end{enumerate}

The gist of the MPC construction is based upon the observation that, whereas the phases i) and ii) are not smoothly connected and correspond to two separate minima of the potential, the phases i) and iii) allow for a smooth transition. As the Higgs boson mass changes sign across the critical boundary that distinguishes these two phases, it is clear that the particle can be naturally light if the scalar potential admits a minimum that lies in the proximity of such a boundary. In terms of the couplings in Eq.~\eqref{eq:LUV}, the MPC scenario is then realised when the couplings of the theory satisfy $\lambda_s\sim\lambda_{hs}\simeq0$, which must hold at a specific scale after the RG-improvement of the scalar potential.      

Different aspects of the MPC scenario were explored in earlier works~\cite{Kannike:2022pva, Kannike:2021iyh, Huitu:2022fcw}, with analyses that only retained the effects of the heaviest scalar in  the loop corrections and in the improvement procedure. One of the targets of the present paper is then to extend this result to include also the contributions of the SM Higgs boson, fermions and gauge bosons. Previous studies of the MPC scenario furthermore focused on a specific area of field space characterised by $h \ll  s$, corresponding to $\lambda_s \ll \lambda_{hs} \ll \lambda_h, \lambda_{hs'}, \lambda_{ss'}$, a particular choice of the theory couplings.\footnote{Note that the $\lambda_{s'}$ coupling does not play a significant role in the symmetry breaking dynamics.} In the present analysis we will also go beyond this assumption by adopting EFT methods designed to cope with scenarios with two or more characteristic energy scales. To understand how these arise naturally in MPC scenarios, consider that the $s'$ VEV is required to vanish in the broken phase of the theory. Consequently, the $s'$ field does not mix with the Higgs boson and the dilaton, and its field-dependent mass is simply given by
\begin{equation}
    m_{s's'}^2(h,s) = \frac{\lambda_{hs'}  h^2 + \lambda_{ss'}  s^2}{2} \,.
    \label{eq:sprimmass}
\end{equation}
The Higgs boson and the dilaton mass are, instead, found upon the diagonalisation of the $2 \times 2 $ mass matrix
\begin{equation}
    \mathcal{M}^2 = \begin{pmatrix} m_{hh}^2( h,  s) && m_{hs}^2( h,  s) \\ m_{hs}^2( h,  s) && m_{ss}^2( h,  s)  \end{pmatrix} = \begin{pmatrix} 3 \lambda_h  h^2 + \frac{1}{2}\lambda_{hs}  s^2 && \lambda_{hs}  h  s \\ \lambda_{hs}  h  s && 3 \lambda_s  s^2 + \frac{1}{2} \lambda_{hs}  h^2 \end{pmatrix}\,,
    \label{eq:MassMatHiggsDilaton}
\end{equation}
with the corresponding eigenvalues given by 
\begin{equation}
    m^2_{1,2}( h,  s) = \frac{1}{2} \left( m_{hh}^2( h,  s) + m_{ss}^2( h,  s) \pm \sqrt{4m_{hs}^2 ( h,  s) + (m_{hh}^2( h,  s) - m_{ss}^2( h,  s))}\right)\,.
    \label{eq:HiggsDilatonEigenV}
\end{equation}
Typical choices of the values of the couplings in Eq.~\eqref{eq:LUV} then yield a mass spectrum characterised by $m_{1,2}^2 \ll m_{s'}^2$, thereby imposing the presence of at least two mass scales onto the theory. Because of the mass hierarchy, earlier studies also simplified the computation of the one-loop correction to the scalar potential by retaining only the contribution sourced by $m_s'$:
\begin{equation}
\label{eq:V1L:msponly}
    V^{(1)}_{m_{s'}}(h,s; \mu) = \frac{1}{64 \pi^2} m_{s's'}^4(h,s) \left( \ln \frac{m_{s's'}^2(h,s)}{\mu^2} - \frac{3}{2} \right)\,.
\end{equation}
In the present work, we go beyond this approximation and investigate the phenomenological implications of different mass spectra within the MPC framework.

\section{EFT treatment}
\label{sec:EFT_treatment}

As a first step toward the application of EFT methods for the study of the MPC scenario, we extend the Lagrangian in Eq.~\eqref{eq:LUV} 
to include the source terms, $J_h$ and $J_s$, for the scalar fields that acquire a VEV.\footnote{These terms need to be added to cancel the tadpole contributions, as we briefly mentioned before.} We also include a cosmological constant, $\Lambda$, that generically results from the matching procedure that we discuss below. For the scalar fields that acquire a VEV, we single out the classical background field components,  $\hat{h}$ and $\hat{s}$, and the corresponding quantum fluctuations, $h_q$ and $s_q$, that obey $\expval{h_q} = \expval{s_q} =  \expval{s_q'}=0$. This is achieved through the replacements: 
\begin{equation}
    h \rightarrow \hat h+h_q, \quad s \rightarrow \hat s + s_q.
    \label{eq:shifts}
\end{equation}
Since the VEV of the $s'$-field vanishes by construction, it is not necessary to consider a dedicated source term in the Lagrangian, nor to split the field into a background and a quantum component. The corresponding replacement is then $s' \rightarrow s'_q$.

The next step is to declare the desired hierarchy in scales, to which the EFT tower must be tailored. To that end, we introduce a dimensionless power counting parameter, $z<1$. There are three possible hierarchies with a heavy $s'$: approximate degeneracy $m_h, m_s \ll m_{s'}$, light dilaton with $m_s \ll m_h \ll m_{s'}$ and heavy dilaton $m_h \ll m_s \ll m_{s'}$. We will study the first two scenarios in some detail, in what follows below, while due to the technical complications associated with the third scenario, we will only resort to qualitative estimates.
\subsection{\texorpdfstring{Approximately degenerate scenario: $m_h,\,m_s \ll m_{s'}$}{Approximately degenerate scenario: mh, ms << ms'}}

In this scenario, the Higgs boson and the dilaton both have masses of the order $\order{100 \text{ GeV}}$ and the DM particle $s'$ has a mass of the order $\order{1 \text{ TeV}}$. Consequently, there are two characteristic mass scales: the heavy scale $\mu_H$ identified with the $s'$ mass as $\mu_H \equiv e^{-1/2}\,m_{s'}$, and the light scale $\mu_L$ that we choose to define in terms of the Higgs boson VEV $\mu_L \equiv  e^{-1/2}\, v$. Here, and throughout the rest of the paper, $v$ and $w$ denote the physical VEVs of the $h$ and $s$ fields, respectively. As usual, these are obtained by minimising the full scalar potential.

We restrict our attention to specific regions of the parameter and field spaces in which the following scaling holds:
\begin{align*}
    \hat{s} &\sim 1, & \hat{h} &\sim z, & \lambda_h(\mu_H) &\sim 1, &  \lambda_{s} (\mu_H) &\sim z^4, &  \lambda_{hs}(\mu_H) &\sim z^2,  
    \\
    \lambda_{s'}(\mu_H) & \sim 1, & \lambda_{hs'}(\mu_H) &\sim  1, & \lambda_{ss'}(\mu_H) &\sim 1, & y_t (\mu_H) &\sim 1, & g_Y(\mu_H) &\sim 1, & g_2(\mu_H) \sim 1,
\end{align*}
where $g_Y$ is the $U(1)_Y$ hypercharge coupling, $g_2$ is the $SU(2)_L$ weak coupling and $y_t$ is the top Yukawa coupling. 

The natural magnitude of our power counting parameter $z$ can be inferred by observing that we count the loop-suppression to be of the order $z^2$, hence
\begin{equation}
    \frac{1}{16 \pi^2} \sim z^{2}.
\end{equation}

The scaling of the field-dependent scalar field masses 
\begin{equation}
W_{xy}^2(\hat h, \hat s)=\frac{\partial^2 \mathcal{L}}{\partial x\,\partial y}\Bigg \rvert_{h= \hat h, s = \hat s}\,,\text{ for }x,y=h,s,s'\,,  
\end{equation}
follows from that of the classical background fields fields and Lagrangian parameters:
\begin{equation}
    W_{s's'}^2 \sim 1, \quad W_{hh}^2 \sim z^2, \quad  W_{ss}^2 \sim z^4, \quad
    W_{hs}^2 \sim z^3. 
\end{equation}
In particular, the requirements $\lambda_{hs'}>0$ and $\lambda_{ss'}>0$, which must be maintained throughout the RGE flow to ensure that the $s'$ field never develops VEV, prevent a possible mass mixing between the involved fields.\footnote{Since the $s'$ field is integrated out below $\mu_H$ to construct an EFT, this condition holds in the region of parameter space where $\lambda_{HS'}(\mu)>0$ and $\lambda_{SS'}(\mu) > 0 $ for $\mu \geq \mu_H$ are both satisfied. Such conditions can always be satisfied for our choice of positive, order 1 values of $\lambda_{HS'}$ and $\lambda_{SS'}$ at $\mu_H$, as can been seen by directly expanding the explicit form of the corresponding $\beta$-functions.} The phenomenological implication is that the natural mass scale of our DM candidate can greatly differ from that shared by the Higgs boson and the dilaton.

The tree-level masses scale as
\begin{equation}
    m_1^2 \sim z^4, \quad m_2^2 \sim z^2, \quad m_{s'}^2 \sim 1, \quad m_Z^2 \sim z^2, \quad m_W^2 \sim z^2, \quad m_t^2 \sim z^2,
\end{equation}
where $m_1^2$ and $m_2^2$ denote the eigenvalues of the Higgs-dilaton mass matrix whose matrix elements are given by $W_{hh}^2$, $W_{hs}^2$ and $W_{ss}^2$, wherein the declared hierarchy ensures that the $s'$ field quanta are much heavier than the remaining particles. This configuration straightforwardly allows to define an EFT valid at energy scales below $m_{s'}$, obtained by integrating out $s'$ from the fundamental theory specified in Eq.~\eqref{eq:LUV}.

\subsubsection{Matching at the heavy scale}

As a first step towards improving the full scalar potential in the EFT formalism, we integrate out the heavy $s'$ field by performing the path integral over all the $s'_q$ configurations. The result is
\begin{equation}
    V^{(1)}_{s'}(\hat{h}, \hat{s}; \mu_H) = \frac{W_{s's'}^4(\hat{h}, \hat{s})}{64 \pi^2}\left( \ln  \frac{W_{s's'}^2(\hat{h}, \hat{s})}{\mu_H^2} - \frac{3}{2} \right),
    \label{eq:V1sp:CW}
\end{equation}
which contributes to the matching of the cosmological constant at the one-loop level. 

Notice that that $V^{(1)}_{s'}$ is evaluated at a fixed (field-independent)  scale $\mu_H = m_s' e^{-1/2} = \sqrt{W^2_{s' s'}}(v,w)\,e^{-1/2}$, where $\expval{h} = v, \expval{s} = w$. Thus, all the couplings contributing to $W_{s',s'}^2$ should also be evaluated at $\mu_H$.

The tree-level matching of the $N$-point amplitudes is trivial up to $\order{z^5}$ because of the absence of terms that are linear in $s'_{q}$, which could contribute as heavy virtual states. Therefore, as explained in more detail in Ref.~\cite{Manohar:2020nzp}, after integrating out the $s'$ field we can reabsorb the shifts of $s$ and $h$ introduced in Eq.~\eqref{eq:shifts}, yielding a low energy Lagrangian written as
\begin{equation}
    \mathcal{L}_{\rm EFT} =  {\tilde \Lambda} + \frac{ \tilde \lambda_{h}}{4} h^4 + \frac{\tilde \lambda_{hs}}{4} h^2 s^2 + \frac{\tilde \lambda_s}{4} s^4 + \mathcal{L}_{\rm top} + \mathcal{L}_{\rm gauge} + \order{z^5},
    \label{eq:Left}
\end{equation}
where the last two terms denote the  SM top quark and gauge boson contributions to the EFT Lagrangian, respectively.\footnote{ This holds between the scales $\mu_H$, where we integrate out $s'_q$ and construct the EFT, and $\mu_L$, where we integrate out $h_q,s_q$. At $\mu_L$ the shift appearing in Eq.~\eqref{eq:shifts} should be reintroduced. The purpose of absorbing the shift in between these two scales is to avoid additional complications in the RG evolution. Also notice that we use the same symbols for the $h$ and $s$ fields in the EFT than in the UV theory, even though these fields are actually different from the UV ones.} 

Therefore, the only non-trivial matching condition involves the cosmological constant in the EFT, which becomes
\begin{equation}
    \tilde{\Lambda}(\mu_H) = V^{(1)}_{s'} (\hat{h}, \hat{s}; \mu_H).
    \label{eq:match:cc}
\end{equation}

\subsubsection{Running to the low energy}

Having constructed the EFT Lagrangian at the scale $\mu_H$, we now need to account for the RG-evolution of its couplings down to the scale $\mu_L$ where the remaining path integral over $s$ and $h$ is taken. The $\beta$-functions for the scalar couplings are given in appendix~\ref{app:a1:betaDegenEFT}.

Since we reabsorbed the shift in the $h,s$ fields over the range $\mu_L < \mu < \mu_H$,  no mass scales appear in the Lagrangian in Eq.~\eqref{eq:Left}. Consequently, $\tilde \Lambda$ does not run between $\mu_L$ and $\mu_H$, so $\tilde{\Lambda}(\mu_L) = \tilde{\Lambda} (\mu_H)$, and the RG evolution only affects the scalar quartics $\tilde \lambda_s$, $ \tilde \lambda_h$, and $ \tilde \lambda_{hs}$, as we will see below.

\subsubsection{\texorpdfstring{Integrating out $h,s$}{Integrating out h, s}}

At this stage we reintroduce the shifts $h \rightarrow \hat{h} + h_q$ and $s\rightarrow \hat{s} + s_q$ to integrate out the $h,s$ fields, obtaining the RG-improved potential:
\begin{equation}
\label{VImproved_Degen}
    V(\hat{h}, \hat{s};\mu_L, \mu_H) = \tilde{\Lambda}(\mu_H) + V^{(0)}(\hat{h}, \hat{s};\mu_L)+ V^{(1)}_{(h,s)}(\hat{h}, \hat{s};\mu_L),
\end{equation}
wherein the tree-level contribution at the low scale is given by
\begin{equation}
    V^{(0)}(\hat{h}, \hat{s};\mu_L) = \frac{\tilde \lambda_h(\mu_L)}{4}\hat{h}^4 + \frac{\tilde \lambda_{hs}(\mu_L)}{4}\hat{h}^2 \hat{s}^2 + \frac{\tilde \lambda_s(\mu_L)}{4}\hat{s}^4,
\end{equation}
and the NLL 1-loop correction at the low scale by\footnote{From hereon, in the rest of the paper, the dependence of the field dependent masses on the background fields is left implicit to simplify the notation.}
\begin{equation}
    V^{(1)}_{(h,s)}(\hat{h}, \hat{s};\mu_L) = \sum_{\substack{i=1, 2,t, W, Z,\\G_0, G_+, G_-}} \frac{\tilde m_i^4(\mu_L)}{64 \pi^2}\left( \ln \frac{\tilde m_i^2(\mu_L)}{\mu_L^2} - \frac{3}{2} \right),
    \label{eq:v1sc1}
\end{equation}
in which $\tilde m_{1,2}^2(\mu_L)$ denote the eigenvalues of the field-dependent mass matrix given by Eqs. \eqref{eq:MassMatHiggsDilaton} and \eqref{eq:HiggsDilatonEigenV} in the EFT.\footnote{Since the Higgs and dilaton masses do not mix with the DM field $s'$ the mass eigenvalues in the EFT are formally given by the same expression as in the full UV theory.} Similarly, we denote the field-dependent masses of the top quark, $W$-and $Z$-bosons by $\tilde m_{t}$, $\tilde m_{W}$ and $\tilde m_{Z}$, respectively:
\begin{equation}
    \tilde m_{t}^2= \frac{1}{2} \tilde{y}_{t}^2 \hat{h}, \quad \tilde m_{Z}^2 = \frac{1}{4} (g_Y^2 + g_2^2) \hat{h}^2, \quad \tilde m_W^2 = \frac{1}{4}g_2^2 \hat{h},
\end{equation}
with the associated couplings, as well as  $\tilde{\lambda}_h$, $\tilde{\lambda}_{hs}$ and $\tilde{\lambda}_s$, all evaluated at $\mu_L$. Formally, Eq.~\eqref{eq:v1sc1} also includes the Goldstone contributions. Their field-dependent masses are given by
\begin{equation}
\tilde m_{G_0}^2 = \tilde m_{G_\pm}^2 =  \tilde\lambda_h(\mu_L) \hat{h}^2 + \frac 12 \tilde\lambda_{hs}(\mu_L) \hat{s}^2.
\end{equation}
However, we point out that in the present MPC construction there is no mixing between the dilaton and the neutral Goldstone boson, and since the Goldstone contributions vanish at the minimum of the potential, we will neglect them in our analysis.

\subsubsection{Tadpole improvement}

A complete computation should also include the evaluation and RG-improvement of the tadpoles $J_{h}$ and $J_{s}$ at the scale $\mu_H$, in the UV theory. However, since in the quasi-degenerate case the tadpoles do not enter into the matching conditions, it is not necessary to detail the exact form of $J_{h}$ and $J_s$ for the purpose of our computation.

\subsubsection{Scaling behaviour under RG running}

In our numerical studies we use Eq.~\eqref{VImproved_Degen} with the restriction $\hat{h} = v \text{ and } \hat{s} = w$ to solve for the quartic couplings as a function of the scalar masses and VEVs. This sets the couplings $\lambda_{hs'}$, $\lambda_{ss'}$ at $\mu_H$, and $\tilde \lambda_{hs}$, $\tilde \lambda_{s}$, $\tilde \lambda_{h}$ at $\mu_L$. To make sure that the chosen scaling of the parameters given by powers of $z$ is consistent, we need to ensure that the resulting hierarchy is maintained by the RG running of the EFT couplings between $\mu_L$ and $\mu_H$, as well as in matching the Higgs and dilaton couplings of the EFT to the corresponding UV couplings. Since in this case the matching is trivial, we only need to consider the effect of RG running.

The only coupling that changes by an appreciable amount during its RG evolution is $\tilde \lambda_h$, which progressively diminishes at higher scales mainly because of the top quark Yukawa coupling. However, as long $\mu_{H}/ \mu_L \lesssim \order{10^2}$, $\tilde \lambda_h$ changes only by less than an order of magnitude along the corresponding RG flow, as can be seen e.g. from Figure~\ref{fig:lam_running}. Hence the power counting introduced above remains valid.

\subsection{\texorpdfstring{Fully hierarchical scenario: $m_s \ll m_h \ll m_{s'}$}{Fully hierarchical scenario: ms << mh << ms'}}

In this subsection, we consider a scenario in which the dilaton mass scale is much lower than that of the Higgs boson, being at most of the order $\order{10 \text{ GeV}}$, while the DM mass is of the order $\order{1 \text{ TeV}}$. 

After choosing a particular scaling of the input couplings at $\mu_H$, we will present the construction of a tower of EFTs, reflecting the three different mass scales of this setup, including the appropriate matching conditions and running between the different scales. Finally, we elucidate on the exact functional form the source terms have to take in order for the tadpoles to be zero, and detail their RG-improvement.

The hierarchy characterising this scenario, $m_s \ll m_{h } \ll m_{s'}$, along with the requirement that the Higgs self-interaction be close to the SM value, can be realised by the following scaling of the couplings appearing in Eq.~\eqref{eq:LUV}: 
\begin{align}
    &\hat{s} \sim 1, \, \hat{h} \sim z^2, \, \, \lambda_h(\mu_H) \sim z^2, \,  \lambda_{s}(\mu_H
    ) \sim z^{10}, \,  \lambda_{hs}(\mu_H) \sim z^{6},  \label{eq:hir_dilaton1} \\ \nonumber
    \lambda_{s'}(\mu_H) \sim z, \, &\lambda_{hs'}(\mu_H) \sim  1, \, \lambda_{ss'}(\mu_H) \sim z^4, \, y_t(\mu_H) \sim z, \, g_Y(\mu_H) \sim z, \, g_2(\mu_H) \sim z.
\end{align}
This allows us to identify three different mass scales: the heavy scale  $\mu_{H} = m_{s'}e^{-1/2}$, where we integrate out the DM candidate $s'$, the intermediate mass scale $\mu_I = v \, e^{-1/2}$, where we integrate out the heavier eigenstate of the Higgs-dilaton system, and the low scale $\mu_{L} \equiv m_{s} e^{-1/2}$, where we integrate out the lighter eigenstate of the Higgs-dilaton system. In this scenario, we associate the heavier eigenstate with the Higgs boson and the lighter eigenstate with the dilaton.

\subsubsection{\texorpdfstring{Integrating out $s'$}{Integrating out s'}}
As mentioned before, the presence of three different mass scales necessitates the construction of a tower of EFTs. As a first step, we consider the EFT obtained by integrating out $s'$ at the heavy scale $\mu_H$, which is now given by

\begin{equation}
    \mathcal{L}^{I}_{\rm EFT} =  {\tilde \Lambda} + \frac{ \tilde \lambda_{h}}{4} h^4 + \frac{\tilde \lambda_{hs}}{4} h^2 s^2 + \frac{\tilde \lambda_s}{4} s^4 +  
    \mathcal{L}_{\rm top} + \mathcal{L}_{\rm gauge} + \order{z^{10}}.
    \label{eq:LeftI}
\end{equation}
where the superscript I emphasises the fact that this EFT holds only between $\mu_H$ and $\mu_I$.
As before, the tree-level matching of the $N$-point amplitudes is trivial up to $\order{z^{10}}$ and the non-trivial matching of the cosmological constant at the one-loop is still level given by Eq.~\eqref{eq:match:cc}. 

\subsubsection{\texorpdfstring{Evolution to the intermediate scale and integrating out the SM particles}{Evolution to the intermediate scale and integrating out the SM particles}}

The next step is to RG evolve the EFT couplings down to the intermediate scale  $\mu_I$, where the Higgs boson, the gauge bosons, and the top quark are to be integrated out in the construction of  another EFT. Since the Higgs field can admit a non-zero VEV, the tree-level matching is non-trivial and must be taken into account.

We observe that the parameter scaling defining the new EFT closely follows that introduced in Eq.~\eqref{eq:hir_dilaton1}. This is because the tree-level matching at the heavy scale is trivial and the imposed hierarchy is preserved by the RG evolution between $\mu_H$ and $\mu_I$ as can be seen from Figure~\ref{fig:lam_running}. Thus, we have

\begin{align}
&\hat{s} \sim 1, \, \hat{h} \sim z^2, \, \td\lambda_h(\mu_I) \sim z^2,  \,  \td \lambda_{s}(\mu_I
    ) \sim z^{10},  \\ \nonumber \td \lambda_{hs}(\mu_I)& \sim z^6,  \, y_t(\mu_I) \sim z, \, g_1(\mu_I) \sim z, \, g_2(\mu_I) \sim z,
\end{align}
and we obtain the following effective Lagrangian  that describes the $s$ degree of freedom\footnote{Note that in all EFTs we use the same symbol for $h$ and $s$ fields for notational simplicity, although in different EFTs these fields are not the same. Instead, to distinguish the different couplings, we denote the couplings in the first EFT (where only $s'$ is integrated out) by a tilde, while the couplings in the second EFT (where the SM particles are also integrated out) are denoted by bar. We will use the same notation also in the next subsection, where we will describe the opposite hierarchy between the dilaton and Higgs.} alone:
\begin{equation}
    \mathcal{L}^{\rm s}_{\rm EFT} = \partial_{\mu} s_q \partial^{\mu} s_q - \bar{\sigma}_{s} s_q - \frac{\bar m_{s}^2}{2} s_q^2 - \frac{\bar \rho_s}{3}s_q^3 - \frac{\bar \lambda_s}{4} s_q^4 - \bar \Lambda + \order{z^{28}}. 
    \label{eq:LeftLightDilaton}
\end{equation}

For the tree-level matching, we first need to find the terms linear in $h_q$. We write 
\begin{equation}
    \mathcal{L}^{I}_{\rm EFT} \supset - X(s_q) h_q, \quad X(s_q) = -J_{h} + \tilde m_{h s}^2 s_q + \frac{1}{2}\tilde \lambda_{hs}\hat{h}s_q^2,
\end{equation}
where $X \sim z^{12}$.
The tree-level matching amplitude is thus given by
\begin{equation}
    \mathcal{L}^{I}_{\rm EFT} \supset  -\frac{1}{2} X \frac{1}{p^2-\td m_{hh}^2(\mu_I)}X = \frac{1}{2} X \frac{1}{\td m_{hh}^2(\mu_I)}X + \frac{1}{2} X \frac{p^2}{\td m_{hh}^4(\mu_I)}X + \order{z^{28}}
\end{equation}
where, neglecting the Higgs-dilaton mixing, we have approximated the field-dependent mass of the Higgs by the $(1,1)$ element of the Higgs-dilaton mass matrix in the EFT, given by Eq.~\eqref{eq:MassMatHiggsDilaton}.
To get the leading order non-trivial effects from matching, we keep only the first term, thus considering matching through $\order{z^{20}}$. The matching procedure then yields:
\begin{equation}
\begin{split}
    \Bar{\sigma}_s(\mu_I) &= -J_{s}(\mu_I) +\frac{J_{h}(\mu_I) \tilde \lambda_{h s}(\mu_I) \hat{h} \hat{s}}{\td m_{hh}^2(\mu_I)},  \\ 
    \bar{m}_s^2(\mu_I) &= \frac{\tilde\lambda_{h s}(\mu_I)}{2}\hat{h}^2 + 3 \tilde \lambda_{s}(\mu_I)\hat{s}^2 + \frac{J_{h}(\mu_I) \tilde \lambda_{hs}(\mu_I) \hat{h} - \tilde \lambda_{h s}^2(\mu_I) \hat{h}^2 \hat{s}^2 }{\td m_{hh}^2(\mu_I)}, 
    \\
    \bar{\rho}_s(\mu_I) &= 3 \tilde \lambda_{s}(\mu_I)\hat{s} - \frac{3}{2}\frac{\tilde \lambda_{hs}^2(\mu_I) }{\td m_{hh}^2(\mu_I)}\hat{h}^2 \hat{s} ,\\
    \bar{\lambda}_s(\mu_I) &=  \tilde \lambda_{s}(\mu_I) - \frac{ \tilde \lambda_{hs}^2(\mu_I) \hat{h}^2}{2 \td m_{hh}^2(\mu_I)}.
\label{eq:EFTMATCHLightDilaton}
\end{split}   
\end{equation}

The matching condition of the cosmological constant at the one-loop level is given by
\begin{equation}
    \Bar \Lambda(\mu_I) = V^{(1)}_{s'}(\hat h, \hat s; \mu_H) + V^{(0)}(\hat h, \hat s;\mu_I) + V^{(1)}_{\rm int}(\hat h, \hat s; \mu_I) - \frac{J_{h}^2(\mu_I)}{\td m_{hh}^2(\mu_I)}, 
    \label{eq:CCMatchLightS}
\end{equation}
where $V_{s'}^{(1)}(\hat{h}, \hat s; \mu_H)$ is given by Eq.~\eqref{eq:V1sp:CW},
\begin{equation}
    V^{(0)}(\hat h, \hat s; \mu_I) = \frac{\tilde \lambda_{h}(\mu_I)}{4} \hat h^4 + \frac{\tilde \lambda_{hs}(\mu_I)}{4} \hat h^2 \hat s^2  +  \frac{\tilde\lambda_s(\mu_I)}{4} \hat s^4 , 
    \label{eq:V0}
\end{equation}
and
\begin{equation}
    V^{(1)}_{\rm int}(\hat{h}, \hat{s}; \mu_I) = \sum_{i=W,Z,t} \frac{\tilde m_i^4(\mu_I)}{64 \pi^4}\left( \ln \left(\frac{\tilde m_i^2(\mu_I)}{\mu_I^2}\right) - c_i \right) + V^{(1)}_{\rm exp}\left(\hat h, \hat s; \mu_I \right),
\end{equation}
and the new term $V_{\rm exp}^{(1)}$ appearing in the potential corresponds to the heaviest eigenvalue of the Higgs-dilaton system, expanded in $z$ through $\order{z^{18}}$:
\begin{equation}
    V^{(1)}_{\rm exp}(\hat h, \hat s; \mu_I) = \frac{\tilde m_{hh}^4(\mu_I)}{64\pi^2}\left( \ln \left( \frac{\tilde m_{hh}^2(\mu_I)}{\mu_I^2}\right) - \frac{3}{2}\right) + \frac{\tilde m_{hs}^4({\mu_I})}{32\pi^2} \left( \ln \left( \frac{\tilde m_{hh}^2(\mu_I)}{\mu_I^2} \right) - 1 \right).
\end{equation}

\subsubsection{\texorpdfstring{RG Evolution to $\mu_L$}{Evolution to muL}}
\label{subsec:lightdilatonrunning}

The RG evolution from $\mu_I$ to $\mu_L$ is governed by the $\beta$-functions given in Appendix \ref{app:a2:betalightdilaton}.  By inspection, we can see that the running procedure will have limited effects from the following reasoning:
\begin{enumerate}
    \item The requirement $m_{s} \ll m_h$ and the hierarchy $w \gtrsim m_{s'} \gg m_h \gg m_s$ entail $\bar \lambda_s (\mu_L) \ll1$. Since the running of $\bar \lambda_s$ is proportional to the same parameter, the RGE flow will not sizeably change its numerical value.
    \item The value of $\bar \rho$ at the matching scale, $\bar \rho_s(\mu_I)$, is suppressed by $\tilde \lambda_{hs}(\mu_I)$. The latter is required to be very small  if the hierarchy between the masses of the Higgs boson and the dilaton is to be maintained. The $\bar \lambda_{s}(\mu_I)$ contribution is also negligible for the reasons given above. The RG evolution of $\bar \rho_s$ is suppressed by $\bar \lambda_s$, so the parameter can only remain close to its small initial value until the scale $\mu_L$ is reached.
    \item Finally, $\bar m_{s}^2$ is required to be very small at $\mu_L$ by construction in the scenario of light dilaton. Moreover, its running, which is proportional to $\bar m_s^2$, $\bar \rho_s^2$ and $\bar \lambda_s$, can thus also have only very limited effects. Consequently, the cosmological constant is effectively frozen to its value at $\mu_I$. 
\end{enumerate}

Since, in this case the system of RGEs has a closed-form analytic solution, we can directly express $\bar \Lambda(\mu_L)$ in terms of $\bar \Lambda (\mu_I), \bar \rho_{s}(\mu_I)$, $\bar m_{s}^2(\mu_I)$, which, using Eq. \eqref{eq:EFTMATCHLightDilaton} can be expressed in terms of the UV theory couplings. We then see that the corrections to the running of $\bar \Lambda$ start at $\order{z^{22}}$:
\begin{equation}
    \bar \Lambda(\mu_L) = \bar \Lambda (\mu_I) + \frac{1}{8} \frac{1}{16\pi^2} \log \left( \frac{\mu_L}{\mu_I} \right) \left( \hat h^2 \lambda_{hs}(\mu_I) + 6 \hat{s}^2 \lambda_s (\mu_I)\right)^2 + \order{z^{24}}.
    \label{eq:ccLightDilatonExpanded}
\end{equation}
Since $\bar \Lambda(\mu_I) \sim z^2$ the effect of running in this EFT is completely negligible. We can repeat the same exercise for $\Bar m_s^2$ that appears in the NLL potential. We obtain
\begin{equation}
   \bar m_{s}^2(\mu_L) = \bar m_{s}^2(\mu_I) + \frac{3}{16\pi^2} \log \left( \frac{\mu_L}{\mu_I} \right) \lambda_{s}(\mu_I) \left( \hat{h}^2 \lambda_{hs}(\mu_I) + 18 \hat{s}^2 \lambda_s(\mu_I) \right) + \order{z^{24}}.
\end{equation}
Thus, the running starts from order $z^{22}$ as was also the case for $\bar \Lambda$. Since $\Tilde m_{s}(\mu_I) \sim z^{10}$, we can neglect the running of $\bar m_{s}$ in the low energy theory. This allows to also obtain a simpler formula for the running of $\bar \Lambda$:
\begin{equation}
    \bar \Lambda(\mu_L) \simeq  \bar \Lambda (\mu_I) + \frac{1}{32\pi^2} \log\left( \frac{\mu_L}{\mu_I} \right) \bar m_{s}^4(\mu_I),  
\end{equation}
consistent with the scaling we found for the running of $\bar \Lambda$ in Eq.~\eqref{eq:ccLightDilatonExpanded}.

\subsubsection{Tadpole improvement}
Finally, we need to ensure that the tadpole conditions are satisfied at the scale $\mu_I$, where we reintroduce the shifts of the scalar field by the corresponding classical background values and demand $\expval{h_q} = 0$, as well as $\expval{s_q} = 0$. To do that, we will follow the techniques outlined in \cite{Manohar:2020nzp}.

The tadpole conditions at $\mu_I$ can be easily found by using the RG evolution of the EFT described by the Lagrangian~\eqref{eq:Left}, where we notice that
\begin{equation}
    \frac{\partial \mathcal{L}^{I}_{\rm EFT}}{\partial J_{h}} = h_q, \quad \text{and} \quad \frac{\partial \mathcal{L}^I_{\rm EFT}}{\partial J_s} = s_q.
\end{equation}
Clearly, these conditions can also be written  using the low-energy EFT containing only the scalar $s$, thereby matching the two theories at the scale $\mu_I$\footnote{Again, we remark that, although we use the same symbol, this is not the same field that appears in $\mathcal{L}^{I}$.}:
\begin{equation}
    \frac{\partial \mathcal{L}_{\rm EFT}^\text{s}}{\partial J_{h}}\Bigg\rvert_{\mu=\mu_I} = h_q, \quad \text{and} \quad \frac{\partial \mathcal{L}_{\rm EFT}^\text{s}}{\partial J_{s}}\Bigg\rvert_{\mu=\mu_I} = s_q,
    \label{eq:Tadpole_EFT}
\end{equation}

Considering Eq.~\eqref{eq:LeftLightDilaton} and the matching conditions, it is clear that the tadpole condition for $J_{s}$ in the EFT is equivalent to that written in the UV theory. Importantly, this condition only affects the behavior of the linear term and leaves untouched the running of masses and vacuum energy of importance for the effective potential computation. 

As for the tadpole condition involving $J_{h}$, by using Eq.~\eqref{eq:Tadpole_EFT} we obtain from the condition $\expval{h_q} = 0$
\footnote{ We introduce the notation $\expval{a}_{\mu}$ to indicate  the expectation value of the operator $a$ evaluated at the scale $t= \frac{1}{16 \pi^2} \ln \left( \frac{\mu}{\mu_I} \right) $.}
\begin{equation}
    J_{h}(\mu_I) = \Tilde \lambda_{h s}(\mu_I)\hat{h}\hat{s}\expval{s_q}_{\mu_I} + \frac{ \Tilde \lambda_{hs}(\mu_I)}{2}\hat{h}\expval{s_q^2}_{\mu_I},
\end{equation}
and setting $\expval{s_q}=0$, as enforced by the second tadpole condition, yields
\begin{equation}
    J_{h}(\mu_I) =  \frac{\tilde \lambda_{hs}(\mu_I)}{2}\hat{h}\expval{s_q^2}_{\mu_I}.
\end{equation}
The expectation value of the composite operator $s_q^2$ at the scale $\mu$, reads 
\begin{equation}
    \expval{s_q^2}_\mu = \frac{1}{16 \pi^2} \bar{m}_s^2 (\mu) \left[ \ln\left( \frac{\bar{m}_s^2(\mu)}{\mu^2}\right) -1 \right].
    \label{eq:chiqatlowscale}
\end{equation}
Hence, if directly evaluated at the scale $\mu_I$, the hierarchy $\bar m^2_s(\mu_I) \ll \mu_I^2$ introduces a large logarithmic correction that needs to be addressed via RG-improvement. For this purpose, we consider the running of $\expval{s_q^2}$ from $\mu_I^2$ to a generic scale $\mu^2$: 
\begin{equation}
    \expval{s_q^2}_{\mu} = \frac{3\bar{m}_s^2(\mu_I)\bar\lambda_s(\mu_I) \eta(t)\left[1 - \eta(t) \right] + \left[ \eta(t) - 1 \right]^2\bar{\rho}_s^2(\mu_I) + 9 \bar{\lambda}_s^2(\mu_I) \expval{s_q^2}_{\mu_I}}{9 \bar{\lambda}^2_s(\mu_I) \eta(t)}, 
    \label{eq:runchiq}
\end{equation}
where,
\begin{equation}
\eta(t) = (1-18 t \bar \lambda_s (\mu_I) + 18 t_I \bar \lambda_
s(\mu_I) )^{1/3},
\end{equation}
and we used $t = \frac{1}{16 \pi^2} \ln \left( \frac{\mu}{\mu_I} \right)$, from which it follows that $t_I  \equiv t(\mu= \mu_I) = 0$. Then, by setting $t=t_L$ in Eq.~\eqref{eq:runchiq} and equating it to Eq.~\eqref{eq:chiqatlowscale} computed for $\mu=\mu_L$, we can solve for $\expval{s_q^2}_{\mu_I}$ and, thus, also for $J_{h}(\mu_I)$.

We note that, in general, the resulting equation for $J_h$ is a complicated transcendental equation, since the EFT couplings initialised at the intermediate scale, corresponding to $t_I=0$,  also depend explicitly on $J_{h}$. In order to have a closed-form solution for $J_{h}$ we thus use
    \begin{equation}
\expval{s_q^2}_{t_L} = 0 
\label{eq:chiqatlowscale2}
\end{equation}
instead of \eqref{eq:chiqatlowscale}, which is well justified for $\td{m}_s(\mu_L) \simeq \mu_L$.

Finally, we argue that the tadpole condition can also be safely neglected. The contribution of $J_{h}(\mu_I)$  enters through the matching conditions at the intermediate scale $\mu_I$ via Eqs.~\eqref{eq:EFTMATCHLightDilaton}
and \eqref{eq:CCMatchLightS}. Similarly to Subsection \ref{subsec:lightdilatonrunning} we can express these contributions in terms of the UV theory couplings at $\mu_I$ and the background fields, providing an easy way to estimate the contribution of these terms to the matching. We obtain for $\bar \Lambda^2 (\mu_I)$ matching, starting at $\order{z^{30}}$, that
\begin{equation}
    \frac{J_{h}^2(\mu_I)}{\Tilde{m}_{hh}^2(\mu_I)} = \frac{\left[ 3 \hat{h}^2 \lambda_{h} \lambda_{hs} - 2\hat{s}^2 \left( \lambda_{hs}^2 - 9 \lambda_h \lambda_s \right) \right]^2 \left( \log \left[\frac{\mu_L}{\mu_I}\right]- \frac{1}{2}  \right)^2 }{4608 \pi^4 \,(\hat{s}^2/\hat{h}^2) \,(\lambda_h^2/\lambda_{hs}) } + \order{z^{32}},
\end{equation}
and for $\bar m_s^2(\mu_I)$ matching, starting at $\order{z^{22}}$, that
\begin{equation}
    \frac{J_{h}(\mu_I) \tilde{\lambda}_{hs}(\mu_I) \hat{h}}{\tilde{m}^2_{hh}(\mu_I)} =  
    \frac{  \left[ 3 \hat{h}^2 \lambda_{h} \lambda_{hs} - 2\hat{s}^2 \left( \lambda_{hs}^2 - 9 \lambda_h \lambda_s \right) \right] \left( \log \left( \frac{\mu_L}{\mu_I} \right) - \frac{1}{2} \right)  }{48 \pi^2 \,(\hat{s}^2/\hat{h}^2) \,(\lambda_h/\lambda_{hs})   }  + \order{z^{24}},
\end{equation}
where all the couplings on the right hand side are assumed to be evaluated at $\mu_I$. Both of these contributions are very small compared to the leading terms in the matching and can thus be safely neglected in numerical studies.

\subsection{\texorpdfstring{Heavy dilaton: $m_h \ll m_s \ll m_{s'}$}{Heavy dilaton: mh << ms << ms'}}

Finally we remark that the MPC construction also allows to consider cases where the Higgs boson is much lighter than the dilaton. 

For a thermal production of DM abundance, the DM mass is bounded from above by the so called Griest-Kamionkowski bound, that follows from considerations of perturbative unitarity\cite{PhysRevLett.64.615}:
\begin{equation}
    m_{s'} \lesssim 100 \text{ TeV}.
\end{equation}
Hence, in the context of the thermal freeze-out scenario, heavy dilaton along with fully hierarchical mass spectra can safely be realised if
\begin{equation}
    m_s \sim \order{1\, \rm TeV}, \quad m_{s'} \sim \order{10\, \rm TeV}.
\end{equation}
Applying the parametrisation in \eqref{eq:SMparametrisation_at_LL} then suggests using the following scaling at $\mu_H$: 
 \begin{align*}
    \hat{s} &\sim 1, & \hat{h} &\sim z^2, & \lambda_h(\mu_H) &\sim z, &  \lambda_{s} (\mu_H) &\sim z^8, &  \lambda_{hs}(\mu_H) &\sim z^4,  
    \\
    \lambda_{s'}(\mu_H) & \sim 1, & \lambda_{hs'}(\mu_H) &\sim  1, & \lambda_{ss'}(\mu_H) &\sim 1, & y_t (\mu_H) &\sim 1, & g_Y(\mu_H) &\sim 1, & g_2(\mu_H) \sim 1.
\end{align*}
Again we can identify three different mass scales: the heavy mass scale $\mu_H \equiv m_{s'}e^{-1/2}$ of the DM candidate, the intermediate scale associated with the dilaton --- the heavier eigenstate of the Higgs-dilaton system, $\mu_I = m_{s}e^{-1/2}$, and finally the low scale $\mu_L = v e^{-1/2}$, which we define in terms of the Higgs VEV.

Although, in contrast to the previous section, performing a complete calculation is complicated by the fact that the RGEs of the low energy theory (where $s$ and $s'$ have been integrated out) do not admit analytical solutions, we can still perform qualitative estimates, as will be shown below. 

\subsubsection{\texorpdfstring{Integrating out $s'$}{Integrating out s'}}

After integrating out the $s'$ field we end up with the same effective Lagrangian as was already given in Eq.\eqref{eq:Left} up to $\order{z^5}$. As before, the tree-level matching of the $N$-point functions is trivial, while the 1-loop matching of the vacuum energy is given by Eq.\eqref{eq:match:cc}.

\subsubsection{Integrating out the dilaton and running to low energy}

Next we integrate out the dilaton to construct another EFT. The corresponding Lagrangian is given by
\begin{equation}
    \mathcal{L}^{\rm h}_{\rm EFT} = \partial_{\mu} h_q \partial^{\mu} h_q - \bar{\sigma}_{h} h_q - \frac{\bar m_{h}^2}{2} h_q^2 - \frac{\bar \rho_h}{3}h_q^3 - \frac{\bar \lambda_h}{4} h_q^4 - \bar \Lambda + \mathcal{L}_{\rm top} + \mathcal{L}_{\rm gauge} + \order{z^{10}}, 
    \label{eq:Left2}
\end{equation}
where the top and gauge contributions have the same functional form as in the UV-theory.

For the tree-level matching, we first need to find the terms linear in $s_q$. We write 
\begin{equation}
    \mathcal{L}^{I}_{\rm EFT} \supset - X(h_q) s_q, \quad X(h_q) = -J_{s} + \tilde m_{h s}^2 h_q + \frac{1}{2}\tilde \lambda_{hs}\hat{s}h_q^2,
\end{equation}
where $X \sim z^{8}$.
The tree-level matching amplitude is thus given by
\begin{equation}
    \mathcal{L}^{I}_{\rm EFT} \supset  -\frac{1}{2} X \frac{1}{p^2-\td m_{ss}^2(\mu_I)}X = \frac{1}{2} X \frac{1}{\td m_{ss}^2(\mu_I)}X + \frac{1}{2} X \frac{p^2}{\td m_{ss}^4(\mu_I)}X + \order{z^{18}},
\end{equation}
where, neglecting the Higgs-dilaton mixing, we have approximated the field-dependent mass of the dilaton by the $(2,2)$ element of the Higgs-dilaton mass matrix in the EFT, given by Eq.~\eqref{eq:MassMatHiggsDilaton}.
To get the leading order non-trivial effects from matching, we keep only the first term, thus considering matching through $\order{z^{14}}$. The matching procedure then yields:
\begin{equation}
\begin{split}
    \Bar{\sigma}_h(\mu_I) &= -J_{h}(\mu_I) +\frac{J_{s}(\mu_I) \tilde \lambda_{h s}(\mu_I) \hat{h} \hat{s}}{\td m_{ss}^2(\mu_I)},  \\ 
    \bar{m}_h^2(\mu_I) &= \frac{\tilde\lambda_{h s}(\mu_I)}{2}\hat{s}^2 + 3 \tilde \lambda_{h}(\mu_I)\hat{h}^2 + \frac{J_{s}(\mu_I) \tilde \lambda_{hs}(\mu_I) \hat{s} - \tilde \lambda_{h s}^2(\mu_I) \hat{h}^2 \hat{s}^2 }{\td m_{ss}^2(\mu_I)}, 
    \\
    \bar{\rho}_h(\mu_I) &= 3 \tilde \lambda_{h}(\mu_I)\hat{h} - \frac{3}{2}\frac{\tilde \lambda_{hs}^2(\mu_I) }{\td m_{ss}^2(\mu_I)}\hat{s}^2 \hat{h} ,\\
    \bar{\lambda}_h(\mu_I) &=  \tilde \lambda_{h}(\mu_I) - \frac{ \tilde \lambda_{hs}^2(\mu_I) \hat{s}^2}{2 \td m_{ss}^2(\mu_I)}.
\label{eq:EFTMATCHheavyS}
\end{split}   
\end{equation}

The matching condition of the cosmological constant at the one-loop level is given by
\begin{equation}
    \Bar \Lambda(\mu_I) = V^{(1)}_{s'}(\hat h, \hat s; \mu_H) + V^{(0)}(\hat h, \hat s;\mu_I) + V_{\rm exp}^{(1)}(\hat{h}, \hat{s}; \mu_I) - \frac{J_{s}^2(\mu_I)}{\td m_{ss}^2(\mu_I)}, 
    \label{eq:CCMathc}
\end{equation}
where $V_{s'}^{(1)}(\hat{h}, \hat s, \mu_H)$ is given by Eq.\eqref{eq:V1sp:CW}, $V^{(0)} (\hat h, \hat s; \mu_I)$ by Eq. \eqref{eq:V0},
and the new term $V_{\rm exp}^{(1)}(\hat{h}, \hat{s}; \mu_I)$ appearing in the potential corresponds to the heaviest eigenvalue of the Higgs-dilaton system, expanded in $z$ through $\order{z^{16}}$:
\begin{equation}
    V^{(1)}_{\rm exp}(\hat h, \hat s; \mu_I) = \frac{\tilde m_{ss}^4(\mu_I)}{64\pi^2}\left( \ln \left( \frac{\tilde m_{ss}^2(\mu_I)}{\mu_I^2}\right) - \frac{3}{2}\right) + \frac{\tilde m_{hs}^4({\mu_I})}{32\pi^2} \left( \ln \left( \frac{\tilde m_{ss}^2(\mu_I)}{\mu_I^2} \right) - 1 \right).
\end{equation}

The crucial difference of the heavy dilaton scenario from the light dilaton one is the presence of gauge bosons and Yukawa in the lowest energy EFT. Consequently, the RGEs of the parameters in  Eq.~\eqref{eq:Left2} cannot be solved analytically, and so it is not possible to provide general estimates like we did in Subsection \ref{subsec:lightdilatonrunning}. Instead one has to go through the entire procedure that we have obtained above, specifying the values of all couplings at $\mu_H$ and then varying the values of $\hat{h}, \hat{s}$ to numerically  compute $\Bar{\Lambda}(\mu_L)$ for each pair of $\hat{h}, \hat{s}$. The specific values of $\hat{h} = v, \hat{s} = w$ that yield the smallest $\bar{\Lambda}$ then correspond to the minimum of the potential, subsequently allowing to compute the mass spectrum.

However, to get a feeling for the importance of running between $\mu_I$ and $\mu_L$ we can perform a simple estimate. The running of $\Bar \Lambda$  is given by
\begin{equation}
    \frac{d \bar \Lambda}{dt} = \beta_{\bar \Lambda}.
\end{equation}
where $t \equiv \frac{1}{16 \pi^2} \log \left( \frac{\mu}{\mu_I} \right) $ and 
\begin{equation}
\beta_{\bar \Lambda} = \frac{1}{2} \left(  \bar m_{h}^4(t) + 6 \bar m_W^4(t) + 3 \bar m_Z^4(t) \right) - 6 \bar m_t^4(t),
\end{equation}
and where we have neglected the Goldstone boson contributions, due to our assumption that we only study the field space around its minimum configuration.
Assuming that all the field-dependent masses stay roughly constant along the RGE flow, we then obtain
\begin{equation}
    \bar \Lambda(t_L) \simeq \beta_{\bar \Lambda}(t_I) t + \bar \Lambda(t_I).
    \label{eq: runningCCHeavyDilaton}
\end{equation}
Then, considering the scaling of the different terms of in the $\beta$-function, and remembering that $t$ is loop suppressed, we see that the change in the cosmological constant due to running from $\mu_I$ to $\mu_L$ approximately scales like $z^{10}$. Notice that the value of $\bar m_{h}^2 (\mu_I)$ is different for different choices of input couplings and dilaton VEV and so this can also result in a small change in the parametrisations given in the Appendix for the case of heavy dilaton. Moreover, going beyond the approximation in Eq.~\eqref{eq: runningCCHeavyDilaton}, by including also the running of the masses, could slightly modify this estimate, especially if the masses are growing at lower energies. Nevertheless, we expect the effect to be small, due to the fact that $\mu_L$ and $\mu_I$ typically differ only by an order of magnitude.

\subsubsection{Tadpole improvement}

For the tadpole improvement, we go through the same procedure as was considered for the light dilaton scenario. In contrast to the former scenario, this time the source term that we are interested in is $J_s$. In analogy to the calculations done before, we now obtain
\begin{equation}
    J_{s}(\mu_I) = \Tilde{\lambda}_{hs}(\mu_I) \hat{h}\hat{s} \expval{h_q}_{\mu_I} + \frac{\Bar{\lambda}_{hs}(\mu_I)}{2}\hat{s}\expval{h_q^2}_{\mu_I}.
\end{equation}
Using the vanishing tadpole condition, $\expval{h_q} = 0$, results in
\begin{equation}
    J_{s}(\mu_I) = \frac{\bar{\lambda}_{hs}}{2}\hat{s}\expval{h_q^2}_{\mu_I}.
\end{equation}
The expectation value of the composite operator $h_q^2$ at the scale $\mu$, that is $\expval{h_q^2}_\mu$, is given by
\begin{equation}
    \expval{h_q^2}_\mu = \frac{1}{16\pi^2} \bar{m}_h^2(\mu) \left[ \log \left( \frac{\bar{m}_h^2}{\mu^2} \right) -1 \right].
\end{equation}
As before, if $\mu_I > \bar m_h(\mu_I)$, large logarithms may appear, calling for RGE improvement. Thus, for proper calculation, one should evaluate $\expval{h_q^2}_{\mu_L}$, where the logarithms are under control, and then run it up to $\mu_I$. Since the running of the composite operator  $\expval{h_q^2}$ depends on the running of the other dimensionful couplings, such as $\bar m_h^2$, we will not go into the details of the full RGE system and simply note from dimensional analysis we expect the running to be proportional to $\bar m^2_h$ and $\bar \rho_h^2$, and assuming that the running of these parameters is relatively mild, we can estimate the change of $\expval{h_q^2}_\mu$ induced by the running to be of the order $z^4$. This is the same order as $\expval{h_q^2}_{\mu_L}$ and so one can expect the running to play a non-negligible role in determining the value of $\expval{h_q^2}_{\mu_I}$. 

Having determined the value of $J_{h}(\mu)$, we finally briefly describe how it affects the overall procedure of computing the effective potential. $J_{h}(\mu_I)$ appears in the matching conditions of $\bar \sigma(\mu_I)$, $\bar m^2(\mu_I)$ and $\bar \Lambda(\mu_I)$. From dimensional analysis, it is clear, that $\bar \sigma$ does not enter in the RGE-s of $\bar m^2$ or $\bar \Lambda$, so we only need to consider the matching of $\bar \Lambda (\mu_I)$ and $\bar m^2 (\mu_I)$. Using the arguments above to argue that $J_{h}(\mu_I) \sim z^6$ we find that the corrections to the matching scale as $z^8$ both for $\bar m_h^2$ and $\bar \Lambda$. Hence the correction from the tadpole matching can probably be neglected for $\bar \Lambda $ that scales as $\bar \Lambda(\mu_I) \sim 1$, while for $\bar m_h^2(\mu_I) \sim z^4$ the effect is less than per mille level, which can generally be neglected as well.

\section{DM phenomenology}
\label{sec:dmpheno}

In order to constrain the properties of the proposed DM candidate, we implemented the model in Eq.~\eqref{eq:LUV} with the FeynRules package~\cite{Alloul:2013bka,Christensen:2008py} and used use the micrOMEGAs code ~\cite{Belanger:2018ccd} to perform a numerical scan of the related parameter space. In addition to requiring our DM candidate to reproduce the observed relic abundance 
$\Omega_{\rm DM} h^{2} = 0.120 \pm 0.001$~\cite{Planck:2018vyg} via thermal freeze-out, we also impose constraints due to the latest direct detection results from the LZ collaboration~\cite{Aalbers2025_LZ4p2ty}, LHC searches targeting the presence of extra scalar states~\cite{Bechtle:2013xfa, Bechtle:2011sb}, as well as perturbative unitarity.

When possible, the obtained numerical results were also checked by means of analytical approximations of the involved expressions. For instance, in the limit $m_{s'}\gg m_s, m_h$, the relativistic DM annihilation cross section is given by \cite{Kannike:2022pva}
\be
\sigma_{\rm ann}  v_{\rm rel} 
\approx\frac{ \lambda_{SS'}^2+4\lambda_{HS'}^2}{64\pi m_{s'}^2}.
\label{eq:sigmavsimple}\ee
For the parametrisation where $\lambda_{h}$ is fixed to its SM value, we obtain
\begin{equation}
    \sigma_{\rm ann}  v_{\rm rel} \approx \frac{4\pi^3 m_{s}^4 }{m_{s'}^6} \left( 9 + 8 \sqrt{1 - \frac{m_h^2}{m_s^2}} - 4 \frac{m_h^2}{m_s^2}\right).
\end{equation}
Across our parameter range of interest, we find that the relic density is very well reproduced if
\begin{equation}
    \sigma_{\rm ann} v_{\rm rel} \approx \frac{1}{M^2}, \quad \text{with } M = 11\,{\rm TeV}.
    \label{eq:sigmav_estimate}
\end{equation}
From this, we can estimate the necessary relation between $m_s$ and $m_{s'}$ to reproduce the DM relic density in the fixed $\lambda_{h}$ parametrisation:
\begin{equation}
    m_{s'} = \left[ 4\pi^3 M^2 m_s^4 \left( 9 + 8 \sqrt{1 - \frac{m_h^2}{m_s^2}} - 4 \frac{m_h^2}{m_s^2}\right) \right]^{1/6}.
\end{equation}
Instead, for the parametrisation where the dilaton VEV is fixed, the general formula is given by

\begin{equation}
    \sigma_{\rm ann} v_{\rm rel} \approx \frac{1}{16 m_{s'}^6 \pi \left( v^2 + w^2 \right)^2} \left[ \frac{(m_{s'}^4 w - v \Delta_{s'})^2}{w^2} + 4 \frac{\left(m_{s'}^4v + w \Delta_{s'}\right)^2 }{v^2} \right],
    \label{eq:Sigmafixedw}
\end{equation}
where 
\begin{equation}
    \Delta_{s'} = \sqrt{\frac{\left(8 \pi^2 m_h^2  (v^2 + w^2) - m_{s'}^4\right) \left( m_{s'}^4 - 8 \pi^2 m_{s}^2  (v^2 + w^2) \right) }{m_{s'}^4 (v^2 + w^2)^2}}.
\end{equation}
Eq.~\eqref{eq:Sigmafixedw} has two important limiting cases. If $m_{s'}^4 \simeq 8\pi^2 m_{h}^2 (v^2+w^2)$, then the cross section becomes effectively independent of $m_s$. Instead, if $m_{s'}^4 \simeq A \times 8\pi^2 m_{s}^2 (v^2 + w^2)$, where A is an $\mathcal{O}(1)$ number, then the cross section scales like $\sigma_{\rm ann} v_{\rm rel} \simeq \left(m_{s}/m_{s'}^2
\right)^2$. The agreement of these formulae with the numeric results are explored in the next section.

The cross section probed in DM direct detection experiments is sourced by the scattering of DM particle on nuclei $N$, mediated by the Higgs boson, which can be described with the effective coupling
\be
 \frac{f_N m_N}{v} h\bar N N,
\ee
where $m_N=0.946$ GeV is the nucleon mass and $f_N\approx 0.3$ is a form factor~\cite{Alarcon:2011zs, Alarcon:2012nr, Cline:2013gha}. Both mass eigenstates $h_{1,2}$, arising from the mass mixing of $h$ with the dilaton $s$, contribute to the spin-independent (SI) direct detection cross section. At the tree level, in the low energy limit, we therefore have
\begin{align}
\sigma_{\rm SI} = \frac{f_N^2 m_N^2 \mu^2}{\pi m_{s'}^2 v^2} 
\left[\frac{\lambda_{h_1 s' s'}}{m_1^2}\cos\theta +\frac{\lambda_{h_2 s' s'}}{m_2^2}\sin\theta\right]^2,
\end{align}
where $\mu=m_{s'}m_N/(m_{s'}+m_N) \simeq m_N$ is the reduced DM/nucleon mass, $\theta$ is the mixing angle that defines the two eigenstates, and $\lambda_{h_i s' s'}$ are the couplings between the eigenstates and DM. For the parameterisation considered in \cite{Kannike:2022pva}, the contributions mediated by $h$ and $s$ simplify, so that the cross section only depends on the DM mass $m_{s'}$:  
\begin{align}
\sigma_{\rm SI}^{\rm prev}\simeq
\frac{64 \pi^3 f_N^2 m_N^4}{m_{s'}^6}.
\label{eq:dd}
\end{align}
In the present paper, we also take into account the mixing between the Higgs and the dilaton, and so strictly speaking the parametric dependence of $\sigma_{\rm SI}$ is more involved than in Eq. \eqref{eq:dd}. 

For the $\lambda_{h}(\mu_L) = \lambda_h^{\rm SM}$ parametrisation, we find, for $m_s \simeq m_h$
\begin{equation}
\sigma_{\rm SI} \simeq \frac{64 \pi^3 f_N^3 m_N^4}{m_{s'}^6} \left[ 1 + \left( \frac{4 \pi m_s}{m_{s'}}\frac{v}{m_{s'}} \right)^2 \left(\frac{4 m_s^2}{m_h^2} \left( 1 + \sqrt{1- \frac{m_h^2}{m_s^2}} \right) - 3 - \sqrt{1- \frac{m_h^2}{m_s^2}} \right)\right],
\end{equation}
and, for $m_s \gg m_h$,
\begin{equation}
    \sigma_{\rm SI} \simeq 
    \frac{64 \pi^3 f_N^3 m_N^4}{m_{s'}^6} \left[ 1 + 2 \pi^2 \left(\frac{m_s}{m_h}\frac{v}{m_{s'}} \frac{m_s}{m_{s'}}\right)^2 \right].
\end{equation}

Thus, we see that as long as we maintain the condition $m_{s'} \gg m_h, m_s$, Eq~\eqref{eq:dd} can be safely used to determine bounds from direct detection. We note that using the more involved formulae modifies the lower bound on $m_{s'}$ mass only by a few percent, at most.

The same approach can also be followed in the scenario in which $w$ is a set input and $\lambda_{h}(\mu_L)$ is allowed to vary. However, in this case the analytical formula is much more involved and so we choose not report it here. Still, we point out that Eq.~\eqref{eq:dd} can be safely used to approximate the spin-independent direct detection cross section also in this scenario, at least in the majority of the parameter space that we are interested in. The only region where this approximation breaks down is around $m_{s'}^4 \simeq 8\pi^2 m_{h}^2 (v^2 + w^2) $. Here, the DM relic abundance can be reproduced for a large range of values for $m_s$ which, in turn, results in sharp jumps of $\sigma_{\rm SI}$ contour near the aforementioned value of $m_{s'}$. However, for the values of $w$ that we are interested in, this feature does not affect the lower bound on $m_{s'}$ from direct detection and so it can be disregarded in our computation. 

\section{Results and discussion}
\label{sec:res}

\begin{figure}[tb]
\begin{center}
  \includegraphics[width=0.33\textwidth]{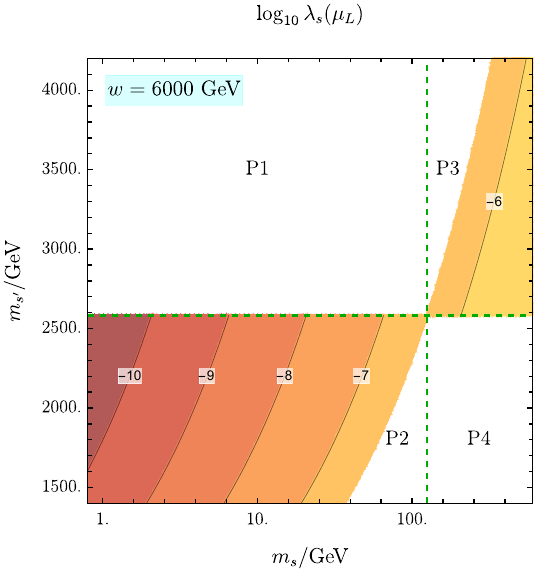}~\includegraphics[width=0.33\textwidth]{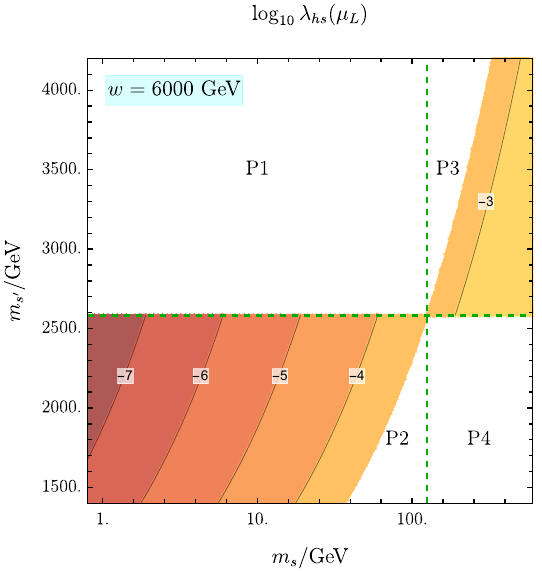}~\includegraphics[width=0.33\textwidth]{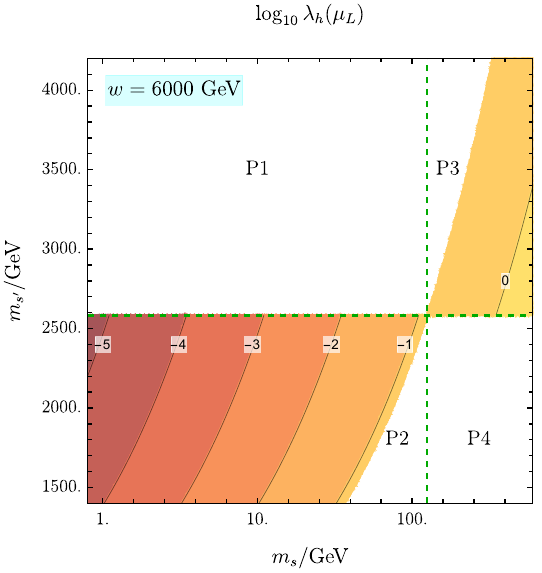} 
  \\
  \vspace{0.67em}
  \includegraphics[width=0.33\textwidth]{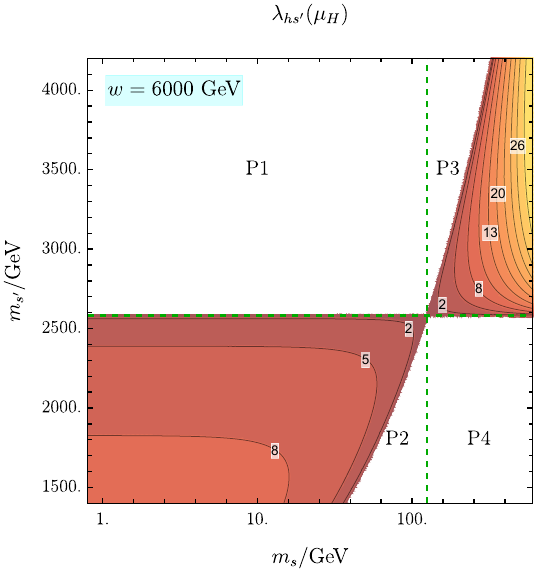}~\includegraphics[width=0.33\textwidth]{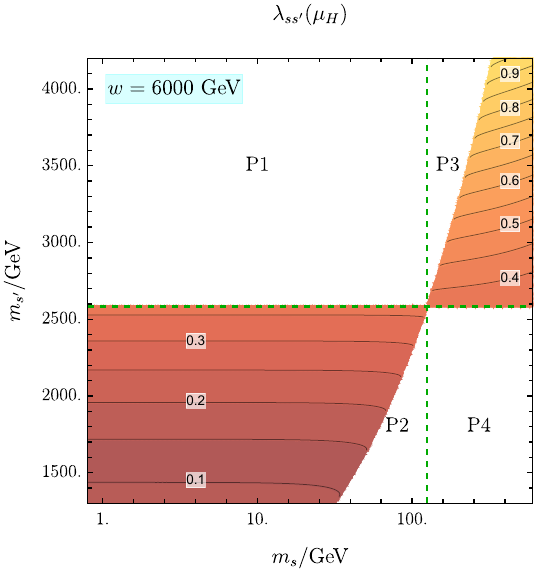}~\includegraphics[width=0.33\textwidth]{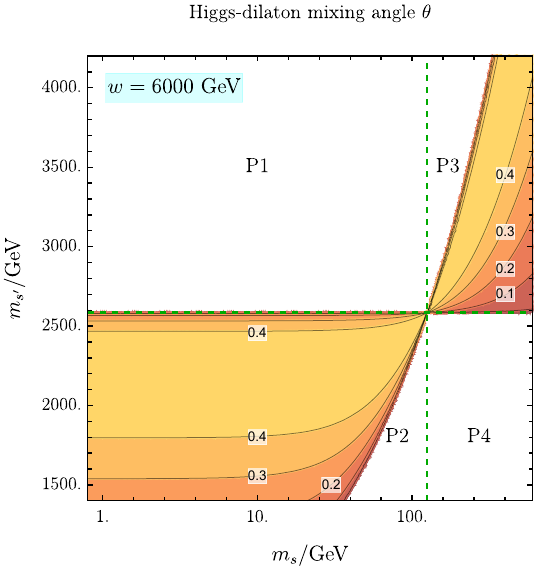} 
\caption{Parametrisation of the quartic couplings and the mixing angle $\theta$ for fixed dilaton VEV, $w = 6000$~GeV.}
\label{fig:params:w:fix}
\end{center}
\end{figure}

We now study the available parameter space of the model, determined by the viability of dynamical symmetry breaking, by the DM relic density and other experimental constraints to be specified below. The parametrisation of the scalar couplings in terms of the physical masses and VEVs is given in Appendix \ref{app:A}. In this section we use only the leading order results in Eqs.~\eqref{eq:parametrisation_at_LL} and ~\eqref{eq:SMparametrisation_at_LL}, the small corrections due to the RG running and higher order terms are discussed in the Appendix, while the effects of the RG-improvement in the light and heavy dilaton scenario are discussed in Section \ref{sec:EFT_treatment} but not included in the numerical study of the parameter space due to their small sizes. 

As a first benchmark, we consider the nearly degenerate case with  dilaton VEV set at $w=6 \text{ TeV}$. The possible values for the scalar couplings and the mixing angle on the $m_s-m_{s'}$ plane are displayed by the coloured contours on Figure~\ref{fig:params:w:fix}. The white region, instead, denotes the area in the $m_s-m_s$ plane where MPC cannot be realised due to the fact that the DM portal couplings become negative.
Upon closer inspection, we can clearly distinguish two different regimes $m_s < m_h$ and $m_s > m_h$, which are divided by the vertical dashed green line at $m_s = 125 \text{ GeV}$. For $m_s < m_h$ the requirement that the DM portal couplings be real throughout the parameter space imposes $m_{s'}^4 < 8 m_h^2 \pi^2 (v^2 + w^2)$ and $m_{s'}^4 > 8 m_s^2 \pi^2 (v^2 + w^2)$, which rule out the regions P1 and P2, respectively. On the other hand, for  $m_h < m_s$, we have $m_{s'}^4 < 8 m_s^2 \pi^2 (v^2 + w^2)$ and $m_{s'}^4 > 8 m_h^2 \pi^2 (v^2 + w^2)$, which rule out the regions P3 and P4, respectively. Thus, for a fixed $w$, we notice an interesting feature of the MPC scenario in the nearly degenerate regime: for $m_s < m_h$ the DM mass is bounded from above, while for $m_h < m_s$ the DM mass is bounded from below. The value of the upper and lower bound is only a function of the dilaton VEV $w$, which is a result of including the Higgs-dilaton mixing in deriving the parametrisation in Eq.~\eqref{eq:parametrisation_at_LL}. This feature  was not observed in the previous studies of MPC scenarios~\cite{Kannike:2022pva}, in which the Higgs-dilaton mixing was ignored in the parametrisation. As one would expect, all the couplings increase with increasing DM and dilaton masses. It can also be seen from Figure~\ref{fig:params:w:fix} that the Higgs-DM portal is the first coupling to approach non-perturbative values which sets the bound for the dilaton mass $m_s \lesssim 300 \text{ GeV}$ for $m_{s'} \simeq \text{a few TeV}$. However, the perturbativity bound becomes less strong at higher values of $m_{s'}$ and so it is still possible to realise the heavy dilaton scenario, with $m_{s} \simeq 1 \text{ TeV}$ if one considers sufficiently high values of $m_{s'}$ and $w$.

Another interesting feature for our chosen value of $w$ is that the scenario $m_s \ll m_h \ll m_{s'}$ leads to extremely small values of Higgs quartic $\lambda_h \lesssim 10^{-3}$. In this case, the Higgs mass is mainly sourced by the loop correction from the $s'$ field. However, we have checked that for higher values of $w$ the Higgs quartic can still take values close to the SM value $\lambda_{h} \simeq \order{0.1}$ for $w \gtrsim 10 \text{ TeV}$.

The values for the Higgs-dilaton mixing angle for $w=6 \text{ TeV}$ are shown on the bottom rightmost panel of Figure~\ref{fig:params:w:fix}. In the case of $m_s \lesssim m_h$, taking $\theta \lesssim 0.3$ forces   $m_{s'}$ to lie either in a narrow bound around $m_{s'} \simeq 2600\, {\rm  GeV}$ or between $1.8 \text{ TeV} <m_{s'} < 2.2 \text{ TeV}$  for $  70 \text{ GeV} \lesssim m_{s} \lesssim 125 \text{ GeV}$. For $m_{s} > m_{h}$, the constraint from the mixing angle can be avoided by increasing the mass of $m_s$, as one would intuitively expect. An additional constraint from $h \rightarrow ss$ applies for $m_s < m_h/2$  which we will discuss shortly.

\begin{figure}[tb]
\begin{center}
  \includegraphics[width=0.33\textwidth]{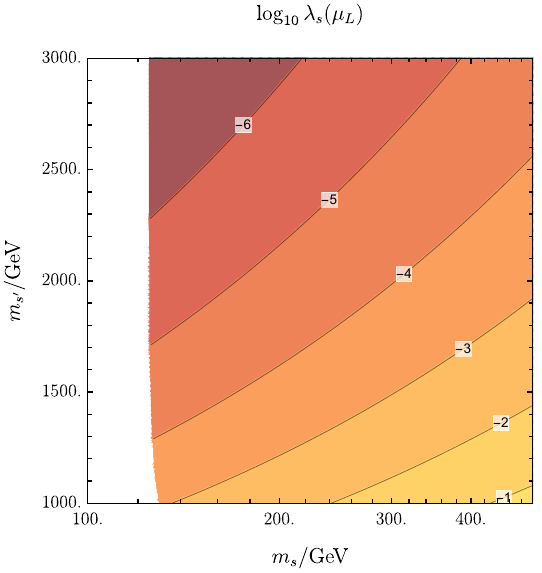}~\includegraphics[width=0.33\textwidth]{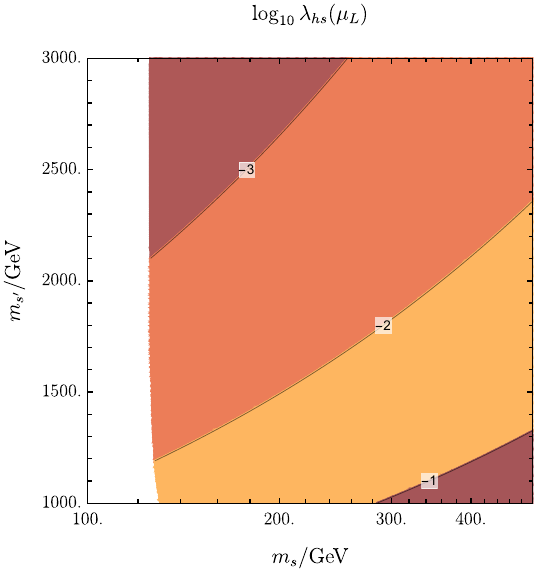}~\includegraphics[width=0.33\textwidth]{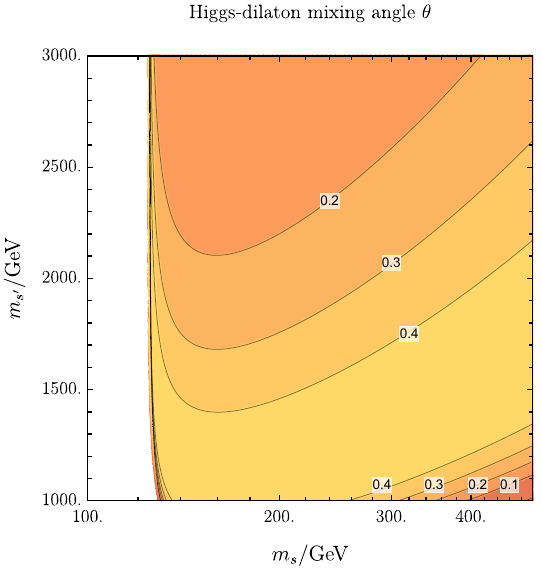} 
  \\
  \vspace{0.67em}
  \includegraphics[width=0.33\textwidth]{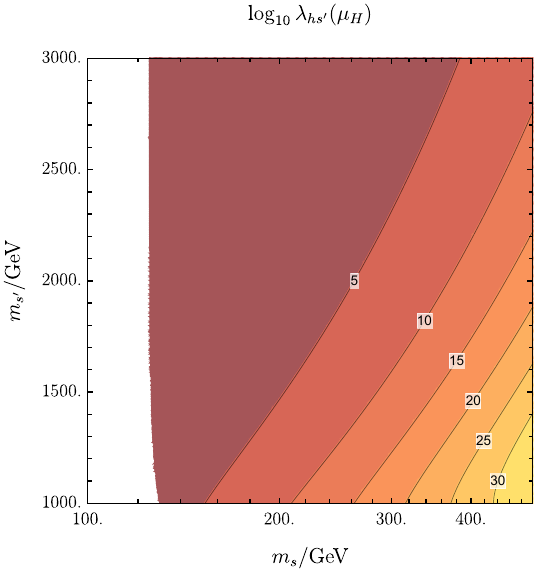}~\includegraphics[width=0.33\textwidth]{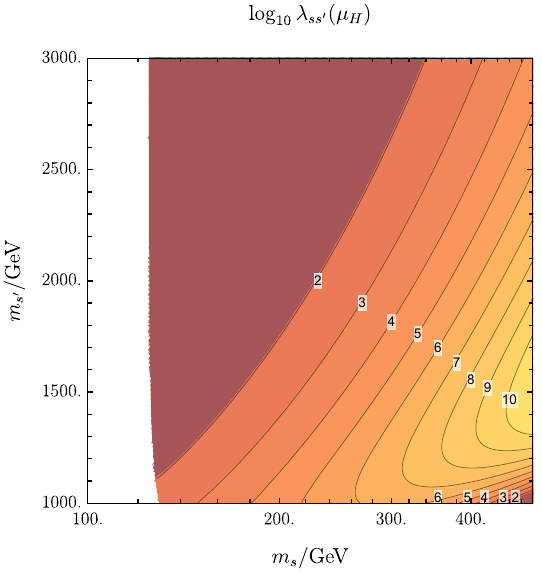}~\includegraphics[width=0.33\textwidth]{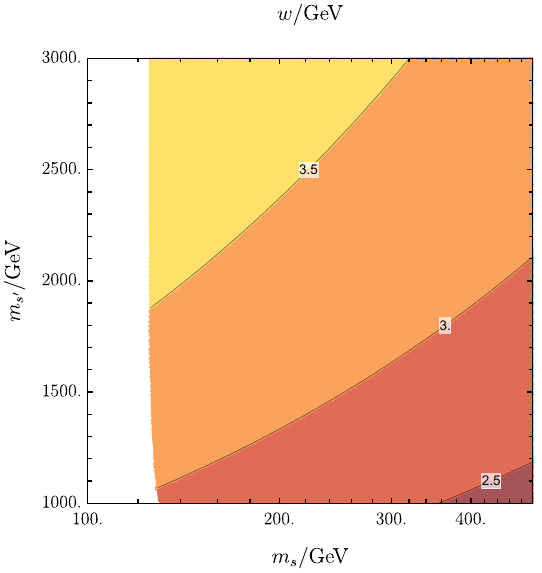} 
\caption{Parametrisation of the quartic couplings and the mixing angle $\theta$ for fixed $\lambda_H = 0.129$.}
\label{fig:params:lH:fix}
\end{center}
\end{figure}
 
Instead of fixing the dilaton VEV $w$, one could alternatively set the Higgs boson self-coupling $\lambda_h$ at its SM value and compute the value of $w$. The analytical form for this parametrisation at the leading order is displayed in Eq.~\eqref{eq:SMparametrisation_at_LL}, and the results obtained with it are displayed in Figure~\ref{fig:params:lH:fix}. 

From Eq.~\eqref{eq:SMparametrisation_at_LL} it can be seen that the requirement that the DM portal couplings to Higgs and the dilaton be real imposes the constraint $m_{s}^2 > \frac{m_h^2 m_{s'}^4}{m_{s'}^4 - 8 \pi^2 v^2 m_h^2}$. For large hierarchies in the scalar mass spectrum, this requires the dilaton to be heavier than the Higgs boson, again confirming our earlier observation that the dilaton cannot be made much lighter than the Higgs boson if the Higgs quartic is fixed close to the SM value. Moreover,  for smaller differences between the DM and the dilaton masses, the upper bound on the dilaton mass becomes stronger, as one would expect. The values of the quartic couplings typically increase with increasing dilaton masses and decreasing DM ones. This is because, for a set value of $m_s$, decreasing $m_{s'}$ reduces the value of the dilaton VEV $w$, as shown in the rightmost panel in the second row of Figure~\ref{fig:params:lH:fix}. Hence, in this case, larger quartic couplings are needed to keep $m_s$ constant. For larger values of $m_s$ and smaller values of $m_{s'}$, the Higgs-DM portal coupling can again take non-perturbative values, similarly to the case with a set value of $w$. The bounds from non-perturbativity start to become relevant from $m_{s'} \lesssim 2 \text{ TeV}$ and $m_{s} \gtrsim 200 \text{ GeV}$.

For larger DM masses, the Higgs-dilaton mixing will always satisfy the bound $\theta \lesssim 0.3$ if a robust upper bound of the DM mass $m_{s'} \lesssim 2500 \text{ GeV}$ is set. For dilaton masses closer to the Higgs boson one the bound relaxes. Furthermore, the bound on the mixing can also be satisfied in the small region in the bottom-right corner of the $m_s - m_{s'}$ plane.

\begin{figure}
    \centering
    \includegraphics[width=0.75\linewidth]{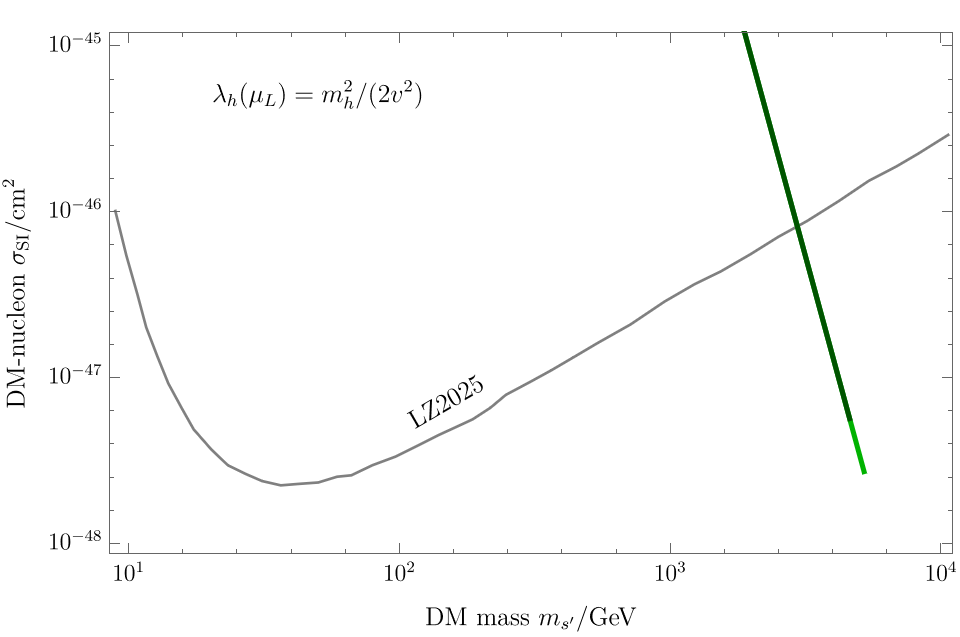}
    \caption{Spin independent direct detection limits for $\lambda_H = \lambda_H^{\rm SM}$ (light and dark green) and for $w = 6000$~GeV (dark green).}
    \label{fig:SI_DD}
\end{figure}

In addition, one can obtain  constraints from the DM relic abundance and direct detection experiments, allowing to connect the dynamical symmetry breaking to DM phenomenology. The direct detection signal is controlled by the SI cross section given by \eqref{eq:dd} as displayed on Figure~\ref{fig:SI_DD}. The latest LZ results \cite{Aalbers2025_LZ4p2ty} then rule out the MPC scenario for $m_{s'} \lesssim 3 \text{ TeV} $, regardless of the value of the dilaton mass.

\begin{figure}[tb]
\begin{center}
  \includegraphics[width=0.49\textwidth]{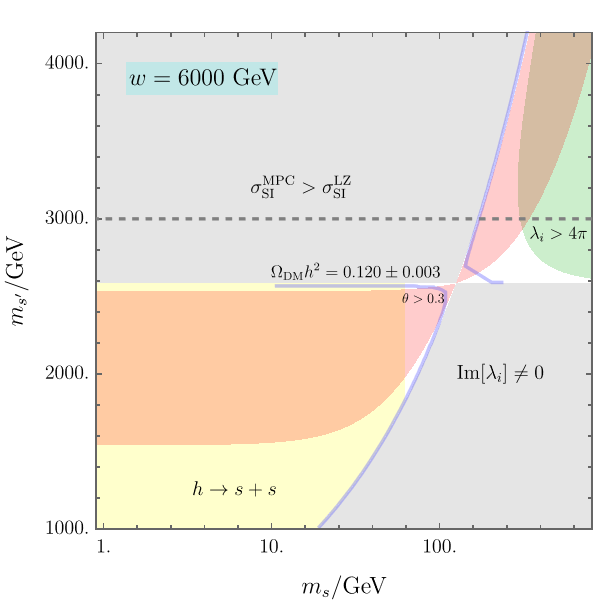}~\includegraphics[width=0.49\textwidth]{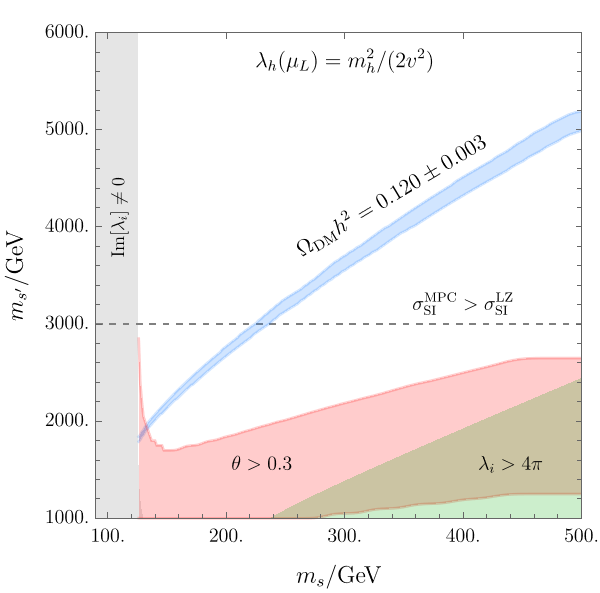}
\caption{Constraint plots for a fixed $w = 6 \text{ TeV}$ (left) and for a fixed $\lambda_h =m_h^2/(2v^2)$ (right).}
\label{fig:Constraintplot}
\end{center}
\end{figure}

To study the connection between DM physics and the MPC scenario in even more detail, we also require that the freeze-out abundance of $s'$ reproduce the observed DM relic density. This is depicted on the two panels of Figure~\ref{fig:Constraintplot}.

In the left panel, we consider a more general scenario, allowing $\lambda_h$ to vary, while fixing the projection of the VEV on the dilaton axis $w=6 \text{ TeV}$. This scenario allows for both a light dilaton with $m_s < m_h/2$ as well as a heavy one $m_s > m_h$. In both cases, the requirement that all scalar couplings be real rules out a large part of otherwise available  parameter space, as discussed before. The bounds from the Higgs-dilaton mixing are strongest in the region where $m_s \sim m_h$ (displayed in light red) as one would intuitively expect, while for $m_s < m_h/2$ there is an extra bound from $h \rightarrow ss$ that requires $\theta < 0.002$ \cite{Robens:2016xkb} that rules out the majority of the parameter space on the left hand side of the $m_s-m_{s'}$ plane aside from small regions around $62.5 \text{ GeV} \lesssim m_{s} \lesssim 125 \text{ GeV}$, where this bound does not apply. 
The upper right corner of the $m_s - m_{s'}$ plane is constrained by the perturbativity of the DM portals, colored in green. However, we have checked that for higher values of $w$ than chosen here, the green region shifts upwards on the $m_{s'}-m_{s}$ plane and thus higher values of the dilaton and DM mass become compatible with the MPC scenario. As we have already discussed, direct detection experiments impose a lower bound $m_{s'} \gtrsim 3 \text{ TeV}$. This is illustrated by the gray dashed line, ruling out the possibility of $m_s < m_h$ for $w = 6 \text{ TeV}$. This is due to the requirement that all scalar couplings be real, imposing $m_{s'}^4 < 8 m_h^2 \pi^2 (v^2 + w^2)$ in this region. However, from this constraint we can also see that, taking high enough VEV $w \gtrsim 8100 \text{ GeV}$, the light dilaton scenario can still be realised.
Finally,  we also compute the DM relic abundance, and identify the parameter space that reproduces the experimentally observed value $\Omega_{\rm DM} h^{2} = 0.120 \pm 0.001$. We can identify two different branches where the observed DM relic density can be reproduced. For both branches we can identify two different curves, a horizontal one centered around $m_{s'} \simeq 2570 \text{ GeV}$ and another piece which grows as $m_{s'} \propto \sqrt{m_{s}}$. The horizontal curves can be explained by the fact that near the boundary between the P1 and P2 (or equivalently P3 and P4) regions on Figure \ref{fig:params:w:fix}, the annihilation cross section in Eq.~\eqref{eq:Sigmafixedw} becomes effectively independent of $m_s$ because one has $\Delta_{s'} \simeq 0$. Thus, the value of $m_{s'}$ that is needed to reproduce the observed DM abundance also becomes effectively independent of the dilaton mass. In this region, the annihilation process is dominated by the Higgs-DM portal coupling. Away from the region near the boundary where the DM portal couplings become imaginary, both branches contain a term that grows like $m_{s'} \propto \sqrt{m_{s}}$. In this region the DM annihilation is also dominated by the Higgs-DM portal.

We notice that the first branch, corresponding to $m_s < m_h$, is ruled out by direct detection. However, this is only an artefact of our chosen $w$. As we have said before, choosing a larger $w$ allows for higher values of $m_{s'}$. In turn, the horizontal line of the first branch near the boundary where the couplings would become imaginary also shifts up. Thus, for high enough value for the dilaton VEV, the observed amount of DM can  still be reproduced if $m_{s} < m_{h}$.

The second branch, where $m_s > m_h$, represents a more viable scenario. In this case, for $m_{s} \gtrsim 200 \text{ GeV}$, the correct relic density can be achieved in agreement with the bounds posed by direct detection and the Higgs-dilaton mixing. In the plot, this is shown by the small gap between the gray and red region.

In the right panel, we consider the scenario in which the Higgs self-coupling has been fixed to its SM tree-level value. As we discussed before, the requirement that all couplings take real values then forces $m_s > m_h$, i.e. the light dilaton scenario cannot be realised. The forbidden region is again highlighted in light gray. Since $m_{s} > m_h$ the constraints from the Higgs invisible decay do not apply, and so from the collider side we only need to consider the bounds on the Higgs-dilaton mixing angle, which rule out the area on the parameter space with $\theta > 0.3$ which is denoted in light red. These bounds force the DM to have mass $m_{s'} \gtrsim 2 \text{ TeV}$. The constraints arising from perturbativity (denoted in light green) largely overlap with the constraints from the Higgs-dilaton mixing, while still introducing additional bounds at low masses for the DM particle. Although not shown here, for $m_s \gtrsim 1 \text{ TeV}$ perturbativity may also rule out regions of parameter space with higher values of DM mass, where the bounds from Higgs-dilaton mixing do not apply. The direct detection again provides an upper bound $m_{s'} > 3 \text{ TeV}$ that is the dominant constraint in the range $m_h <m_s < 600 \text{ GeV}$. For higher values of dilaton mass, the requirement of perturbativity becomes the dominant bound instead. The slice of the parameter space that can reproduce the observed abundance of DM is denoted in light blue. Combining the relic density analysis with the direct detection constraints we see that $s'$ can indeed be a valid DM candidate if $m_{s} \gtrsim 2\, m_{h}$.

Finally, to identify the energy scale until which our construction remains valid, we identify the scale of the Landau pole, by running the couplings of the UV-theory up, until
\begin{equation}
    \lambda_{i}(\Lambda) \geq 4\pi, \text{ or } g_{j}(\Lambda) \geq 4\pi, \text{ or } y_{t}(\Lambda) \geq 4\pi,
\end{equation}
where $\lambda_{i}$ denotes all the scalar couplings of the UV theory, $g_{j}$ all the gauge couplings and $y_t$ the top Yukawa coupling. For concreteness, we only consider the scenario in which all the SM couplings are fixed to their SM values at $\mu_L = v\, e^{-1/2}$, the $\lambda_{hs}$ and $\lambda_{s}$ couplings are also initialised at $\mu_L$ and $\lambda_{hs'}$, $\lambda_{ss'}$ at $\mu_H = e^{-1/2} m_{s'}$ according to the parametrisation in Eq.~\eqref{eq:SMparametrisation_at_LL}. The DM self-coupling $\lambda_{s'}$ that does not have an effect on the symmetry breaking dynamics is set to $\lambda_{s'} (\mu_H) = 0.1$. We then plot the scale of the Landau pole as a function of $m_{s'}$, choosing $m_s$ by requiring that the correct DM relic density be reproduced. We see that for low DM masses $m_{s'} < 2 \text{ TeV}$ the scale of the Landau pole is a few orders of magnitude below the GUT-scale. However, such values of DM mass are ruled out by direct detection. For $m_{s'} \gtrsim 3 \text{ TeV}$, allowing to escape the direct detection bound, we observe that the theory breaks down at rather low scales. This happens mainly due to the high values of $\lambda_{ss'}$ and $\lambda_{hs'}$ at $\mu_{H}$ which are needed for the symmetry breaking mechanism to work. However, in all cases we observe that the scale of the Landau pole is higher than the highest scale in our theory (which in this case is the projection of the VEV on the dilaton axis) and so the MPSC scenario remains consistent for all values of $m_{s'}$.

\begin{figure}
    \centering
    \includegraphics[width=0.6\linewidth]{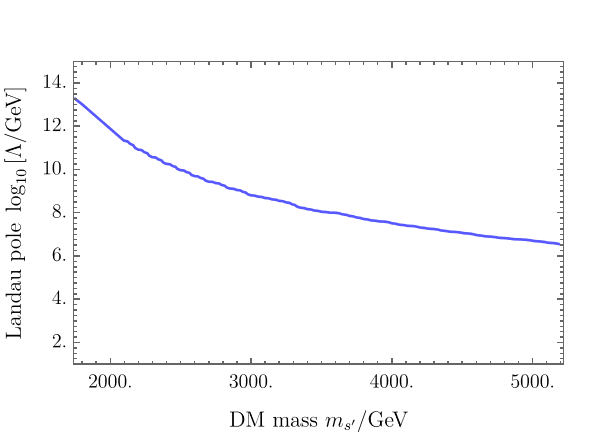}
    \caption{The Landau pole scale at which the theory becomes nonperturbative for the case where $\lambda_H$ has been fixed to the SM value at $\mu_L$. $m_s$ has been chosen in such a way that the correct relic density is reproduced within $3 \sigma$ limits.}
    \label{fig:Landaupole}
\end{figure}

\section{Conclusions}
\label{sec:concl}

We studied dynamical symmetry breaking in a SM extension containing two new scalar degrees of freedom: the dilaton, $s$, and a DM candidate $s'$. Electroweak symmetry breaking is driven by large portal couplings to DM while the scalar couplings are arranged such that the breaking takes place close to the boundary with a phase where EW symmetry is not broken, thereby implementing the MPC scenario. As a consequence, not only the dilaton mass, but also the Higgs mass is naturally light, being suppressed by loop corrections responsible for generating the electroweak scale. 

We show that the hierarchy between DM and other scalar masses cannot be ignored and, therefore, EFT must be used to obtain correct results. Besides that, we have further improved on previous studies by taking into account the Higgs-dilaton mixing, which is important when their masses are of similar size. 

Calculations are simplified by the fact that the DM field does not acquire a VEV. Because of that, for the simplest scenario of nearly degenerate Higgs-dilaton mass spectrum, its one-loop contribution to the effective potential does not run, so we can, effectively, initialise DM couplings at the electroweak scale together with the Higgs and dilaton couplings. 

For a very light dilaton mass, the EFT procedure is technically more complicated, but numerically nothing changes: the running in the low energy theory is controlled by the couplings that give the dilaton its mass and is thus strongly suppressed. The dilaton couplings have to be small to produce the large VEV acquired by this field.

We also briefly considered the opposite hierarchy where the Higgs boson is lighter than the dilaton. In this case, the dilaton portal to the DM is comparable to the Higgs-DM portal, in contrast to the light dilaton scenario where it had to be small. Similarly, however, the running effects in the EFT containing only the lightest state are negligible, since they are suppressed by the hierachy between the Higgs and the dilaton. The effects of properly RG-improved source were estimated to be small for the same reason.

Therefore, we believe that in the MPC scenario, constructing the EFT without the DM particle is generally enough to capture the leading effects. However, our analytical results may still prove useful for more complicated scenarios where considering a tower of EFTs is necessary: for example, in scenarios where the mass spectra span many orders of magnitude or for different scaling of dimensionless couplings of the model.

Thanks to the RGE improvement, we keep the NLL contributions under theoretical control, and so we can also consider one-loop contributions from the Higgs, dilaton, gauge, and Yukawa couplings. Since $m_s'$ is an order of magnitude larger than all other masses, these contributions to the scalar couplings are at the percent level. They are crucial, however, to reproduce the measured values of the Higgs boson VEV and mass.

Reproducing the observed dark matter relic density selects definite ranges for the dilaton and dark matter masses, as seen from Figure \ref{fig:Constraintplot}. The direct detection bound from the LZ experiment constrains the DM mass to be above 3~TeV as seen in Figure~\ref{fig:SI_DD}.


\acknowledgments 
We would like to thank Aneesh V. Manohar for very helpful correspondence on the tadpole improvement. This work was supported by the Estonian Research Council grants PRG803, TEM-TA23, RVTT3, RVTT7,  TARISTU24-TK10, TARISTU24-TK3, and the Centre of Excellence program TK202 ``Fundamental Universe''. KM was supported by the Estonian Research Council personal grant PUTJD1256.

\appendix

\section{Parametrisations of the scalar couplings}
\label{app:A}

\subsection{Parametrisation in terms of masses and VEVs}
\label{app:A1}

In this section we report the parametrisation for the scalar couplings in terms of the physical masses and VEVs of the involved fields. For concreteness, we consider the parametrisation for the approximately degenerate case $m_{s} \approx m_h$. As is shown in the main text, the light dilaton case $m_s \ll m_h$ and the heavy dilaton case $m_{s} \gg m_h$  exhibit only small deviations from the results obtained with this parametrisation. 

Similar formulae were also given in earlier works on MPC and DM phenomenology \cite{Kannike:2022pva}, which differ from the present study in several important aspects.

\begin{enumerate}
    \item Previously, the hierarchy between the light scale $\mu_L = v\, e^{-1/2}$ and the heavy scale $\mu_H = e^{-1/2}\, m_{s'}$ was not taken into account. As we shall show in next section of the Appendix, running between $\mu_L$ and $\mu_H$ usually has a considerable effect on $\lambda_h$ due to the SM contributions, that were also previously not included. In contrast the running of $\lambda_{hs}$, $\lambda_s$ couplings can be mostly neglected, while $\lambda_{hs'}, \lambda_{ss'}$,  being integrated out, do not run at all.
    \item The one-loop logarithmic corrections from the light scalars --- the dilaton and the Higgs --- as well as from the top quark and gauge bosons were also neglected. In this study we include these corrections in an effective description and quantify their magnitude.
    \item Finally, in this work we also include the effects of mass mixing involving the Higgs and the dilaton. This results in differences from the parametrisation of~\cite{Kannike:2022pva} even if the hierarchy of scales and the logarithmic corrections from the light scalars, the gauge bosons and the top quark are not included.
\end{enumerate}

Taking these novel aspects into account, we can write for the couplings at the high scale where we initialise the model parameters,\footnote{Recall that for the degenerate case, the tree level matching between the UV theory and the effective theory is trivial, and so we use the same symbol for the coupling in the EFT and in the UV theory.}

\begin{equation}
    \lambda_{i}(\mu_H) = \lambda_{i}^{\rm LL}(\mu_L) + \Delta \lambda_{i}^{\rm NLL}(\mu_L) + \Delta \lambda_{i}^{\rm run}, \quad \lambda_{i} = \lambda_{hs}, \lambda_{s}, \lambda_{h},
\end{equation}
and 
\begin{equation}
    \lambda_{i}(\mu_H) = \lambda_{i}^{\rm LL}(\mu_H) + \Delta \lambda_{i}^{\rm NLL}(\mu_H), \quad \lambda_{i} = \lambda_{hs'}, \lambda_{ss'},
\end{equation}
with $\lambda^{\rm LL}$ denoting the solutions that include, in addition to the tree-level potential, only the one-loop potential from the DM particle $s'$. In the equation above $\Delta_{i}^{\rm NLL}$ are the corrections brought by the one-loop contributions sourced by the SM particles and the dilaton, while $\Delta \lambda_{i}^{\rm run}$ incorporates the effects of running from $\mu_L$ to $\mu_H$.

Explicitly, we have
\begin{align}
    \nonumber
    \lambda_{s}^{\rm LL}(\mu_L) &= 4 \pi^2 \frac{m_h^2}{m_{s'}^2}\frac{m_s^2}{m_{s'}^2} \frac{v^2}{w^2}, \quad \lambda_{h}^{\rm LL}(\mu_L) = 4\pi^2 \frac{m_h^2}{m_{s'}^2} \frac{m_s^2}{m_{s'}^2} \frac{w^2}{v^2}, \quad \lambda_{hs}^{\rm LL}(\mu_L) = - 8 \pi^2 \frac{m_h^2}{m_{s'}^2}\frac{m_s^2}{m_{s'}^2}, \\
    \lambda_{hs'}^{\rm LL}(\mu_H) &= 2 \frac{m_{s'}^2}{v^2 + w^2} + 2 \frac{w}{v} \Delta_{s'}, \quad \lambda_{ss'}^{\rm LL}(\mu_H) =  2 \frac{m_{s'}^2}{v^2 + w^2} - 2\frac{v}{w}\Delta_{s'},
\label{eq:parametrisation_at_LL}
\end{align}
with 
\begin{equation}
    \Delta_{s'} = \sqrt{\frac{\left(8 \pi^2 m_h^2  (v^2 + w^2) - m_{s'}^4\right) \left( m_{s'}^4 - 8 \pi^2 m_{s}^2  (v^2 + w^2) \right) }{m_{s'}^4 (v^2 + w^2)^2}},
\end{equation}
where $v \equiv \expval{h}$ and $w \equiv \expval{s}$.
For typical values of $m_{s'}$, the changes in $\lambda_{hs}$ and $\lambda_{s}$ due to running are at the level of $10 \%$, while $\lambda_{h}$ changes at most by $50 \%$ for typical values of $m_{s'}$.

The $\Delta \lambda^{\rm NLL}_{i}$ terms, on the other hand, entail corrections at most of the percent level. They are crucial, however, to correctly reproduce the Higgs VEV and the Higgs and dilaton masses, which would otherwise deviate by about 5\% from the experimental values. Approximate analytical expressions for the $\Delta \lambda^{\rm NLL}_{i}$ terms can be easily obtained by expanding the effective potential to the linear order in $\Delta \lambda^{\rm NLL}_{i}$. Because they are extremely lengthy, we do not give them explicitly.

\subsection{Parametrisation with the SM Higgs quartic}

\subsubsection{Parametrisation from the leading-log terms}

In the parametrisation given by the previous subsection, the Higgs quartic $\lambda_h(\mu_L)$ was allowed to vary. In this subsection, instead, we present an alternative parametrisation where $\lambda_{h}(\mu_L)$ is set to its SM value, while $w$ must be computed from the input parameters. Again, we have

\begin{equation}
    \lambda_{i}(\mu_H) = \lambda_{i}^{\rm LL}(\mu_L) + \Delta \lambda_{i}^{\rm NLL}(\mu_L) + \Delta \lambda_{i}^{\rm run}, \quad \lambda_{i} = \lambda_{hs}, \lambda_{s},
\end{equation}
\begin{equation}
    \lambda_{i}(\mu_H) = \lambda_{i}^{\rm LL}(\mu_H) + \Delta \lambda_{i}^{\rm NLL}(\mu_H), \quad \lambda_{i} = \lambda_{hs'}, \lambda_{ss'},
\end{equation}
and
\begin{equation}
    w = w^{\rm LL} + \Delta w^{\rm NLL}
\end{equation}
with $\lambda_i^{\rm LL}$ and $w^{LL}$ denoting the solutions that include, in addition to the tree-level potential, only the one-loop potential from the DM particle $s'$. The $\Delta \lambda_{i}^{\rm NLL}$ and $\Delta w^{\rm NLL}$ corrections due to the one-loop contributions are sourced by the SM particles and the dilaton, while $\Delta \lambda_{i}^{\rm run}$ incorporates the effects of running from $\mu_L$ to $\mu_H$.

For the LL contributions that we use in the numerical studies, we have

\begin{align}
    \nonumber
    w^{\rm LL} &= \frac{1}{2 \sqrt{2}\pi} \frac{m_s'}{m_s}m_s'   
    & 
    \lambda_{s}^{\rm LL}(\mu_L) &= 32\pi^4 \frac{m_h^2}{m_{s'}^2} \frac{m_s^4}{m_{s'}^4} \frac{v^2}{m_{s'}^2}, 
    & 
    \lambda_{hs}^{\rm LL}(\mu_L) &= - 8 \pi^2 \frac{m_h^2}{m_{s'}^2}\frac{m_s^2}{m_{s'}^2}, 
    \\
    \lambda_{hs'}^{\rm LL}(\mu_H) &= 16\pi^2 \frac{m_{s} (m_s m_{s'}^2 + \Delta) }{m_{s'}^4 + 8 \pi^2 m_s^2 v^2 }, \quad
    \lambda_{ss'}^{\rm LL}(\mu_H) =  -16 \pi^2  \frac{m_s^2 (m_{s'}^6 + 8 \pi^2 m_s v^2 \Delta)}{m_{s'}^8 + 8 \pi^2 m_s^2 m_{s'}^4 v^2  }, \span \span \span \span
\label{eq:SMparametrisation_at_LL}
\end{align}
wherein,
\begin{equation}
    \Delta = \sqrt{m_s^2 m_{s'}^4 - m_h^2 \left( m_{s'}^4 + 8 \pi^2 m_s^2 v^2 \right)}.
\end{equation}

\subsubsection{Next-to-leading-log corrections}

Contrary to the parametrisation which sets a value for $w$, the input scheme that sets $\lambda_{h}(\mu_L)$ to a desired value allows to express the NLL effects with rather compact formulae which we present here.
To calculate the shifts induced by the next-to-leading log (NLL) corrections, we expand the NLL potential up to $O(z^4)$, capturing the dominant terms of each field. We find that in order to keep VEVs set at $v=246 \text{ GeV}$ and $w=w^{\text{LL}}$, the parametrisation must be modified by including the following NLL corrections
\begin{align}
\label{eq:Delta_lambda}
    \Delta \lambda_{hs}^{\rm NLL} &\simeq \frac{m_s^2 v^2}{16 m_s'^4} 
    \left( \xi^{\rm top} + \xi^{\rm gauge} \right), \\ \nonumber
    \Delta \lambda_{s}^{\rm NLL} &\simeq \frac{m_s^4 v^4}{4096 m_{s'}^8 \pi^2} \left[ -262 144 \pi^6 \lambda_H - \left(\xi^{\rm top} + \xi^{\rm gauge} \right)\left( 512(2\pi^2 + \lambda_H ) + \xi^{\rm gauge} + \xi^{\rm top} \right) \ \right],
\end{align}
wherein
\begin{align}
\label{eq:chitop}
    \xi^{\rm top} &= 48 yt^4 \log \left( \frac{y_t^2}{2} \right),
\\
\label{eq:chigauge}
    \xi^{\rm gauge} &= 3\left( g_2^2 + g_Y^2\right)^2 \log \left( \frac{4}{g_2^2 + g_Y^2} \right) - 2g_Y^2 \left( 2g_2^2 + g_Y^2\right) - 6 g_2^4 \left[\log \left( \frac{g_2^2}{4} \right) + 1\right],
\end{align}
where the couplings in Eqs. \eqref{eq:Delta_lambda}, \eqref{eq:chitop} and \eqref{eq:chigauge} are understood to be evaluated at $\mu_L$. Notice that the parametric dependence of the NLL corrections on $m_{s}, m_{s'}$ is the same as that of the LL contributions, and so these corrections always take the same value as $m_s$ and $m_{s'}$ are varied. By numerical analysis we obtain
\begin{equation}
    \frac{\Delta \lambda_{hs}^{\rm NLL}}{\lambda_{hs}^{\rm LL}} \simeq 0.09, \quad \frac{\Delta \lambda_{s}^{\rm NLL}}{\lambda_s^{\rm LL}} \simeq 0.1.
\end{equation}
In addition, to keep the Higgs mass at its experimentally observed value, we found numerically that the $\lambda_H$ coupling must be increased by approximately $5 \%$. With these modifications, the observed values of Higgs mass and VEV can be reproduced without altering the DM couplings. Thus, the dilaton and DM masses that receive the dominant corrections from these couplings also remain unchanged.

\subsubsection{Corrections from running}

\begin{figure}
    \centering
    \includegraphics[width=0.76\linewidth]{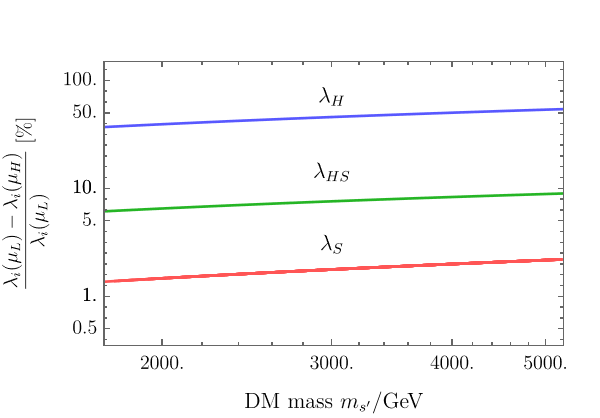}
    \caption{Relative change (in percentage) of $\lambda_{i}$ between the high scale $\mu_H$ where DM decouples, and the low scale $\mu_L$ for the case where $\lambda_H$ has been fixed to the SM value at $\mu_L$ (solid line). 
    $m_s$ has been chosen in such a way that the correct relic density is reproduced within $3 \sigma$ limits.}
    \label{fig:lam_running}
\end{figure}

In previous studies \cite{Kannike:2022pva} all couplings were initialised at the heavy scale $\mu_H$. In this work we have argued that for the approximately degenerate scenario $m_h \simeq m_s$, the couplings involving only the light degrees of freedom should be initialised near the electroweak scale instead. Thus, to obtain a parametrisation at $\mu_H$ one has to take into account the running between $\mu_H$ and $\mu_L$ induced by the light degrees of freedom. This is shown in Figure~\ref{fig:lam_running}. We notice that the running of $\lambda_h$ definitely cannot be neglected, being of the order of $50 \%$. Changes in the Higgs-dilaton portal and the dilaton self-coupling due to running are, instead, below $10 \%$.

\section{Running in the effective theories}

We define the $\beta$-functions by
\begin{equation}
    \frac{\text{d} \lambda}{\text{d} t} = \beta_{\lambda}, \quad t = \frac{1}{16 \pi^2} \ln \frac{\mu}{\mu_0}
\end{equation}
with $\mu_0$ denoting an arbitrary reference scale where we start the running.

\subsection{EFT for the approximately degenerate case}
\label{app:a1:betaDegenEFT}

The $\beta$-functions for the Lagrangian in Eq.~\eqref{eq:Left} are given by
\begin{align}
\beta_{\tilde\lambda_{s}} = 2 \left( 9 \tilde\lambda_s^2 + \tilde\lambda_{hs}^2 \right), \quad \beta_{\tilde\lambda_{hs}} = 2\tilde\lambda_{hs}(\tilde{Z}_h + 3\tilde\lambda_s + 2\tilde\lambda_{hs}), \\
\beta_{\tilde\lambda_{h}} = 4 \tilde{Z}_h \tilde \lambda_h - 6 \tilde y_t^4 + 9 \frac{\tilde g_2^4}{8} + \frac{9}{40}\tilde g_Y^4 + \frac{3}{4}\tilde g_2^2 \tilde g_Y^2 + \frac{1}{2}\tilde \lambda_{hs}^2,
\end{align}
where
\begin{equation}
    \tilde{Z}_{h} = 3 \tilde y_t^2 - \frac{9}{4}\tilde g_2^2 - \frac{3}{4}\tilde g_Y^2 + 6 \tilde \lambda_h,
\end{equation}
and for the top quark and gauge bosons:
\begin{equation}
    \beta_{\tilde g_{1}} = \frac{41}{6}\tilde g_Y^3, \quad \beta_{\tilde g_{2}} = -\frac{19}{6}\tilde g_2^3, \quad \beta_{\tilde g_{3}} = -7\tilde g_3^3, \quad \beta_{\tilde y_t} = \tilde y_t \left( \frac{9}{2}\tilde y_t^2 - 8 \tilde g_3^2 - \frac{9}{4}\tilde g_2^2 - \frac{17}{12} \tilde g_Y^2 \right).
\end{equation}

\subsection{\texorpdfstring{EFT for the dilaton with $m_s \ll m_h \ll m_{s'}$}{EFT for the dilaton with ms << mh << ms'}}
\label{app:a2:betalightdilaton}
The $\beta$-functions for the EFT containing only the dilaton field, summarised by the Lagrangian in Eq.~\eqref{eq:Left2}, are given by
\begin{equation}
 \beta_{\bar{\Lambda}_s} = \frac{1}{2}\bar{m}^4_s, \quad \beta_{\bar{\sigma}_s} = 2\bar{\rho}_s \bar{m}_s^2, \quad \beta_{\bar{m}_s^2} = 6 \bar{\lambda}_s\bar{m}_s^2  + 4 \bar{\rho}_s^2, \quad \beta_{\bar{\rho}_s} = 18 \bar{\lambda}_s\bar{\rho}_s, \quad \beta_{\bar{\lambda}_s} = 18 \bar{\lambda}_s^2.
\end{equation}



\bibliography{CWEFT}
\end{document}